\documentclass[aps,prb,twocolumn,amsmath,amssymb,groupedaddress,longbibliography
]{revtex4-2}

\usepackage{graphicx}
\usepackage{dcolumn}
\usepackage{bm}
\usepackage{multirow}
\usepackage{braket}
\usepackage{color}
\usepackage{ulem}
\usepackage{times}
\usepackage{booktabs}

\begin{document}

\title{
Spin-orbital-momentum locking under odd-parity magnetic quadrupole ordering
}

\author{Satoru Hayami$^1$ and Hiroaki Kusunose$^2$}
\affiliation{
$^1$Department of Applied Physics, the University of Tokyo, Tokyo 113-8656, Japan \\
$^2$Department of Physics, Meiji University, Kawasaki 214-8571, Japan 
 }

\begin{abstract}
Odd-parity magnetic and magnetic toroidal multipoles in the absence of both spatial-inversion and time-reversal symmetries are sources of multiferroic and nonreciprocal optical phenomena. 
We investigate electronic states caused by an emergent odd-parity magnetic quadrupole (MQ) as a representative example of magnetic odd-parity multipoles. 
It is shown that spontaneous ordering of the MQ leads to an antisymmetric spin-orbital polarization in momentum space, which corresponds to a spin-orbital momentum locking at each wave vector. 
By symmetry argument, we show that the orbital or sublattice degree of freedom is indispensable to give rise to the spin-orbital momentum locking. 
We demonstrate how the electronic band structures are modulated by the MQ ordering in the three-orbital system, in which the MQ is activated by the spin-dependent hybridization between the orbitals with different spatial parities. 
The spin-orbital momentum locking is related with the microscopic origin of cross-correlated phenomena, e.g., the magnetic-field-induced symmetric and antisymmetric spin polarization in the band structure, the current-induced distortion, and the magnetoelectric effect. 
We also discuss similar spin-orbital momentum locking in antiferromagnet where the MQ degree of freedom is activated through the antiferromagnetic spin structure in a sublattice system. 
\end{abstract}

\maketitle

\section{Introduction}
\label{sec:Introduction}

The interplay among internal degrees of freedom in electrons, such as charge, spin, and orbital, gives rise to unconventional physical phenomena in the strongly-correlated electron system.
The concept of atomic-scale multipole can describe not only electronic order parameters but also resultant physical phenomena in a unified way~\cite{Kusunose_JPSJ.77.064710,kuramoto2009multipole,Santini_RevModPhys.81.807, Hayami_PhysRevB.98.165110,suzuki2018first}. 
There are four types of multipoles according to space and time inversion properties: electric, magnetic, electric toroidal, and magnetic toroidal multipoles~\cite{dubovik1990toroid,hayami2018microscopic,kusunose2020complete}. 
Moreover, such an atomic-scale multipoles have been generalized so as to describe anisotropic charge and spin distributions over multiple sites, which are denoted as a cluster multipole~\cite{hayami2016emergent,Suzuki_PhysRevB.95.094406,Suzuki_PhysRevB.99.174407,Huyen_PhysRevB.100.094426,Huebsch_PhysRevX.11.011031,hayami2021essential} and a bond multipole~\cite{Hayami_PhysRevLett.122.147602,Hayami_PhysRevB.102.144441}. 
The generalization of the concept of multipole 
is referred to as an augmented multipole, which opens a new direction of cross-correlated (multiferroic) physical phenomena in antiferromagnets, such as the anomalous Hall and Nernst effects in  Mn$_3$Sn~\cite{nakatsuji2015large,ikhlas2017large,Suzuki_PhysRevB.95.094406,Liu_PhysRevLett.119.087202,Zhang_PhysRevB.95.075128,higo2018anomalous}, current-induced magnetization in UNi$_4$B~\cite{saito2018evidence,Hayami_PhysRevB.90.024432,Hayami_1742-6596-592-1-012101,Yanagisawa_PhysRevLett.126.157201} and Ce$_3$TiBi$_5$~\cite{shinozaki2020magnetoelectric,shinozaki2020study}, and nonreciprocal magnon excitations in $\alpha$-Cu$_2$V$_2$O$_7$~\cite{Gitgeatpong_PhysRevB.92.024423,Gitgeatpong_PhysRevLett.119.047201,Hayami_doi:10.7566/JPSJ.85.053705,Gitgeatpong_PhysRevB.95.245119,piyawongwatthana2021formation}. 

The active multipole moments in real space are related with the electronic band structures in momentum space~\cite{Hayami_PhysRevB.98.165110,Watanabe_PhysRevB.98.245129,hayami2019momentum,Hayami_PhysRevB.101.220403,Hayami_PhysRevB.102.144441}. 
In other words, the band deformations and spin splittings at each wave vector are ascribed to the active multipole moments. 
For example, a lowering of the symmetry in the band structure caused by spontaneous electronic orderings, such as the Pomeranchuk instability~\cite{pomeranchuk1959sov,Yamase_JPSJ.69.332,Halboth_PhysRevLett.85.5162} and electronic nematic state~\cite{chuang2010nematic,Goto_JPSJ.80.073702,yoshizawa2012structural} corresponds to the appearance of a particular type of active electric quadrupole. 
Another example is an antisymmetric band-bottom shift without both spatial-inversion and time-reversal symmetries, which is accounted for by the emergent magnetic toroidal dipole moment~\cite{volkov1981macroscopic,kopaev2009toroidal,Yanase_JPSJ.83.014703,Hayami_PhysRevB.90.024432,Hayami_doi:10.7566/JPSJ.84.064717,Matsumoto_PhysRevB.101.224419}. 

The correspondence between the multipole and the band deformation is classified according to the space and time inversion symmetries~\cite{Hayami_PhysRevB.98.165110}; the even(odd)-rank electric (magnetic toroidal) multipole leads to the symmetric (antisymmetric) band deformation and the odd-rank magnetic (electric) multipole and even-rank magnetic toroidal (electric toroidal) multipole induce the symmetric (antisymmetric) spin splittings with respect to the wave vector. 
A systematic classification of the band structure based on multipoles can lead to a further intriguing situation, such as the symmetric/antisymmetric spin splittings even without the spin-orbit coupling~\cite{naka2019spin,hayami2019momentum,Hayami2020b,Hayami_PhysRevB.101.220403,Hayami_PhysRevB.102.144441,Yuan_PhysRevB.102.014422,egorov2021colossal,Yuan_PhysRevMaterials.5.014409,Naka_PhysRevB.103.125114,yuan2021strong}. 

In the present study, we focus on the role of the magnetic quadrupole (MQ) on the electronic structures in momentum space.  
The MQ is characterized as a rank-2 axial tensor with time-reversal odd among the magnetic multipoles. 
As this is a higher-rank multipole of the magnetic dipole, it is defined as the spatial distributions of the magnetic moments, such as the spin and orbital angular momenta.   
A typical example to exhibit the MQ is the antiferromagnetic ordering without the spatial inversion symmetry, which has been discussed in the context of multiferroic materials in magnetic insulators~\cite{Fiebig0022-3727-38-8-R01,cheong2007multiferroics,tokura2014multiferroics,spaldin2019advances,bhowal2021revealing}, such as Cr$_2$O$_3$~\cite{FolenPhysRevLett.6.607,astrov1961magnetoelectric,izuyama1963theory,Shitade_PhysRevB.100.224416,Meier_PhysRevX.9.011011}, Ba(TiO)Cu$_4$(PO$_4$)$_4$~\cite{kimura2016magnetodielectric,Kato_PhysRevLett.118.107601,kimura2020imaging}, Co$_4$Nb$_2$O$_9$~\cite{Khanh_PhysRevB.93.075117,Khanh_PhysRevB.96.094434,Yanagi2017,Yanagi_PhysRevB.97.020404,matsumoto2017symmetry}, and KOsO$_4$~\cite{Song_PhysRevB.90.245117,Yamaura_PhysRevB.99.155113,Hayami_PhysRevB.97.024414}. 
Meanwhile, such ordering has recently been discussed in magnetic metals~\cite{thole2018magnetoelectric,Gao_PhysRevB.98.060402,thole2020concepts,Sato_PhysRevB.103.054416}, such as BaMn$_2$As$_2$~\cite{Watanabe_PhysRevB.96.064432,Shitade_PhysRevB.98.020407}, as it could exhibit intriguing current-induced magnetization and distortion. 

In generalization of the studies on the MQ from insulators to metals, there is a natural question how the MQ affects microscopically the electronic band structure and induces related physical phenomena. 
There is a missing link between the electronic band modulations and the odd-parity magnetic multipoles because the latter cannot be constructed by a simple product between the wave vectors and spin degrees of freedom~\cite{Hayami_PhysRevB.98.165110}. 
To answer this naive question, we investigate the characteristic feature of the electronic band structure by the formation of the MQ ordering based on the simplest multi-orbital model. 
We show that there is a nontrivial spin-orbital entanglement once the MQ ordering occurs; the electric quadrupole polarization defined by the product of spin and orbital angular momenta appears with a dependence on the wave vector in an antisymmetric way. 
By analogy with the spin momentum locking in noncentrosymmetric metals~\cite{hsieh2009tunable}, we refer to the effective spin-orbital entanglement as ``spin-orbital momentum locking". 
The spin-orbital momentum locking becomes the origin of the antisymmetric spin polarization under the magnetic field. 
Moreover, we find that the spin-orbital momentum locking provides a deep understanding of conductive phenomena in magnetic metals, such as the current-induced distortion. 
We also show that a similar spin-orbital momentum locking is realized in the antiferromagnetic ordering by taking into account the sublattice degree of freedom instead of the orbital one. 

The rest of this paper is organized as follows. 
In Sec.~\ref{sec:Magnetic quadrupole}, after introducing the expressions of the MQ in real space, we present them in momentum space, which indicates an origin of the spin-orbital momentum locking. 
We discuss the realization of such spin-orbital momentum locking by considering the multi-orbital and sublattice systems with the active MQ degree of freedom in Sec.~\ref{sec:Active magnetic-quadrupole system}. 
We relate the spin-orbital momentum locking to physical phenomena, such as the current-induced distortion and magnetoelectric effect, under the MQ ordering from the analysis of the linear response theory. 
Section~\ref{sec:Summary} is devoted to the summary.

\section{Magnetic quadrupole}
\label{sec:Magnetic quadrupole}
In this section, we show the expressions of the MQ in real and momentum spaces on the basis of multipole expansion and group theory.
After briefly reviewing the expressions in real-space in Sec.~\ref{sec:Expressions in real space}, we describe the band modulations under the MQ in Sec.~\ref{sec:Spin-orbital-momentum locking}. 
It is shown that the active MQ gives rise to the spin-orbital momentum locking at each wave vector. 

\subsection{Expressions in real space}
\label{sec:Expressions in real space}

\begin{figure}[t!]
\begin{center}
\includegraphics[width=0.8 \hsize]{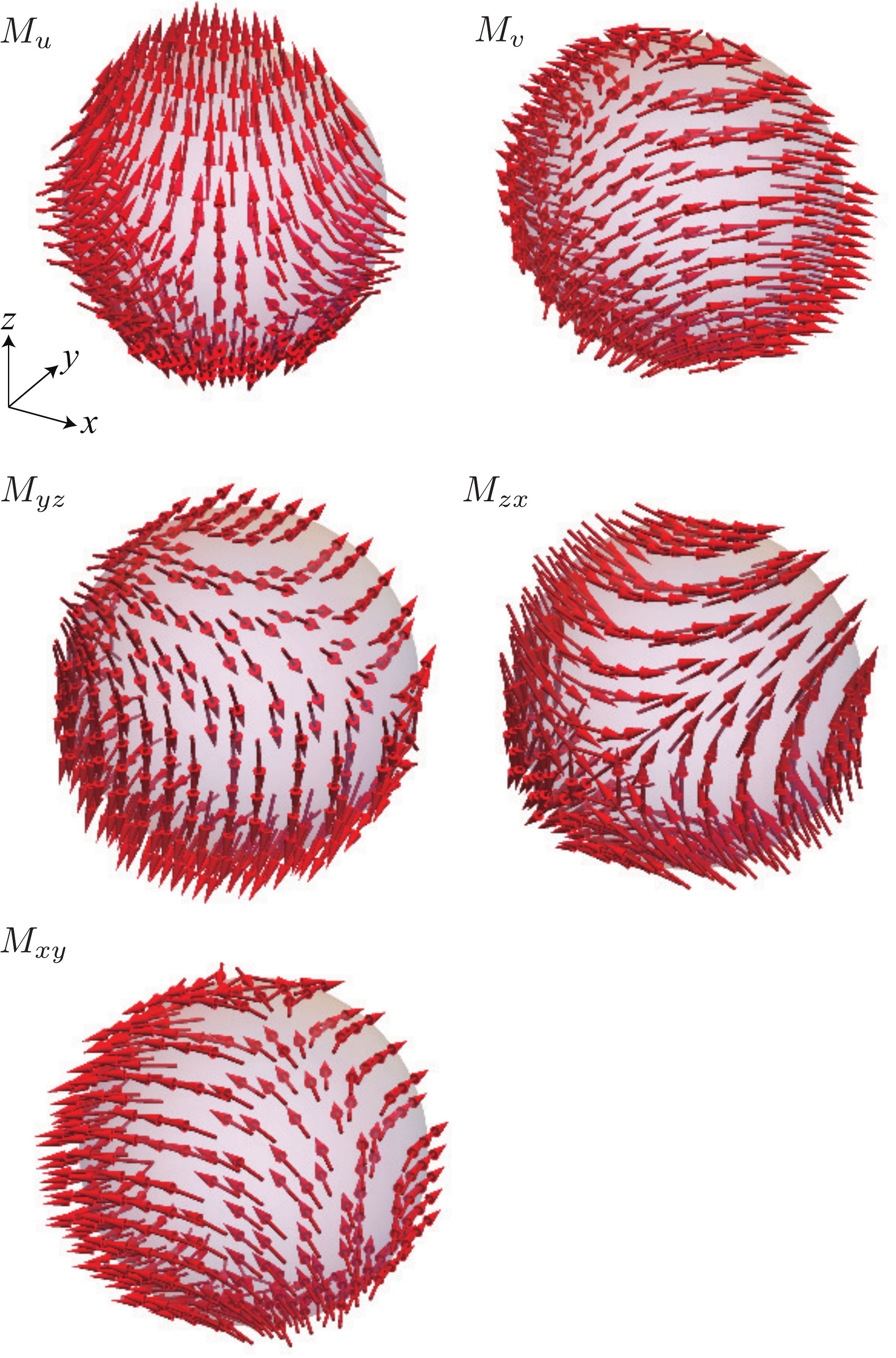} 
\caption{
\label{Fig:MQ}
Schematic pictures of the spin configuration of the MQs, $\{M_u, M_v, M_{yz}, M_{zx}, M_{xy}\}$, in Eqs.~(\ref{eq:Mu})-(\ref{eq:Mxy}) onto the sphere.
The arrows represent the direction of the magnetic moments. 
}
\end{center}
\end{figure}

We start by reviewing the MQ in real space, whose expression is obtained in the second order of the multipole expansion for the vector potential~\cite{LandauLifshitz198001,Spaldin_0953-8984-20-43-434203,hayami2018microscopic}. 
The MQ is characterized by the rank-2 axial tensor with five components $\{ M_u, M_v, M_{yz}, M_{zx}, M_{xy}\}$ with $u=3z^2-r^2$ and $v=x^2-y^2$, which has the odd parities for both space and time inversion operations.
The expressions are given by 
\begin{align}
\label{eq:Mu}
M_u&=2z m_z-x m_x-y m_y, \\
\label{eq:Mv}
M_v&=\sqrt{3}(x m_x-y m_y), \\
\label{eq:Myz}
M_{yz}&=\sqrt{3}(z m_y+ y m_z), \\
\label{eq:Mzx}
M_{zx}&=\sqrt{3}(x m_z+ z m_x), \\
\label{eq:Mxy}
M_{xy}&=\sqrt{3}(y m_x+ x m_y), 
\end{align}
where $\bm{r}=(x,y,z)$ is the position vector and $\bm{m}=(m_x, m_y, m_z)$ is the magnetic moment consisting of the dimensionless orbital and spin angular-momentum operators $\bm{l}$ and $\bm{\sigma}/2$, as $\bm{m}=2\bm{l}/3+\bm{\sigma}$~\cite{hayami2018microscopic}.  
One can confirm that the sign of the MQs in Eqs.~(\ref{eq:Mu})-(\ref{eq:Mxy}) is reversed by the spatial-inversion or time-reversal operations, $\mathcal{P}$ and $\mathcal{T}$, as $\mathcal{P}$ ($\mathcal{T}$) reverses the sign of $\bm{r}$ ($\bm{m}$). 
Meanwhile, the expressions in Eqs.~(\ref{eq:Mu})-(\ref{eq:Mxy}) are invariant under the $\mathcal{PT}$ operation. 
Although such space-time inversion properties are common to the magnetic toroidal dipole proportional to $\bm{r} \times \bm{m}$, they are distinguished from the rotational property: the MQ is the rank-2 axial tensor and the magnetic toroidal dipole is the rank-1 polar tensor.
The real-space spin configurations of each MQ component projected onto the sphere are schematically shown in Fig.~\ref{Fig:MQ}.

\subsection{Spin-orbital-momentum locking}
\label{sec:Spin-orbital-momentum locking}

\begin{table}[htb!]
\caption{
The correspondence between the multipoles and the momentum in terms of the space-time inversion symmetries. 
Even (Odd) in parentheses represents the even(odd)-rank multipoles.
The upper panel shows the correspondence in the single-orbital system, while the lower panel shows that in the multi-orbital system where $\mu, \nu=x,y,z$. 
}
\label{tab_kmp}
\centering
\begin{tabular}{ccccc}
\hline\hline
momentum & $\mathcal{P}$& $\mathcal{T}$ & multipoles  \\
\hline 
$\bm{k}^{2n}$ & $+1$ & $+1$ & electric (even)  \\
$\bm{k}^{2n+1}$ & $-1$ & $-1$ & magnetic toroidal (odd)   \\
$\bm{k}^{2n}\bm{m}$ & $+1$ & $-1$ & magnetic (odd) or magnetic toroidal (even)  \\
$\bm{k}^{2n+1}\bm{m}$ & $-1$ & $+1$ & electric (odd) or  electric toroidal (even)  \\
\hline
$\bm{k}^{2n}l_\mu \sigma_\nu$ & $+1$ & $+1$ & electric (even) or electric toroidal (odd)  \\
$\bm{k}^{2n+1}l_\mu \sigma_\nu$ & $-1$ & $-1$ & magnetic (even) or magnetic toroidal (odd)  \\
\hline \hline
\end{tabular}
\end{table}

In contrast to the real-space expressions of the MQ in Sec.~\ref{sec:Expressions in real space}, it is nontrivial to deduce their momentum-space expressions owing to the opposite time-reversal parity between $\bm{r}$ and the wave vector $\bm{k}$. 
Indeed, it is impossible to construct the counterparts of Eqs.~(\ref{eq:Mu})-(\ref{eq:Mxy}) by replacing $\bm{r}$ with $\bm{k}$. 
In fact, any contractions of the time-reversal odd polar vector $\bm{k}$ and the time-reversal odd axial vector $\bm{m}$ lead to the odd-rank electric multipoles or the even-rank electric toroidal multipoles with time-reversal even rather than the time-reversal odd rank-2 axial tensor, MQ. 
We summarize the correspondence between the multipoles and their $\bm{k}$ dependence in terms of the space-time inversion symmetries in the upper panel of Table~\ref{tab_kmp}. 
Thus, it is concluded that the appearance of the real-space MQ does not affect the electronic band structure within the product between $\bm{k}$ and $\bm{m}$. 

Such a situation is resolved by taking into account the product of two angular momenta, $\bm{l}$ and $\bm{\sigma}$. 
Since both $\bm{l}$ and $\bm{\sigma}$ are the same rank-1 axial tensors but they are independent with each other, their contraction expresses the rank-0 and rank-2 polar tensor $\mathcal{Q}_{\mu\nu}=(l_{\mu} \sigma_{\nu}+l_{\nu} \sigma_{\mu})/\sqrt{2}$ for $\mu,\nu=x,y,z$ and rank-1 axial tensor $\bm{\mathcal{G}}=\bm{l}\times \bm{\sigma}$ with time-reversal even. 
By taking a further contraction between $\bm{k}$ and $\bm{\mathcal{G}}$ or $\mathcal{Q}_{\mu\nu}$, one can construct the rank-2 axial tensor corresponding to the MQ in momentum space. 
In the case of the contraction between $\bm{k}$ and $\bm{\mathcal{G}}$, the MQ in momentum space is expressed as 
\begin{align}
\label{eq:Mu_k1}
M^{\rm (I)}_u(\bm{k})&=2 \mathcal{G}_z k_z - \mathcal{G}_x k_x- \mathcal{G}_y k_y, \\
\label{eq:Mv_k1}
M^{\rm (I)}_v(\bm{k})&=\sqrt{3}( \mathcal{G}_x k_x -  \mathcal{G}_y k_y), \\
\label{eq:Myz_k1}
M^{\rm (I)}_{yz}(\bm{k})&=\sqrt{3}( \mathcal{G}_y k_z+  \mathcal{G}_z k_y), \\
\label{eq:Mzx_k1}
M^{\rm (I)}_{zx}(\bm{k})&=\sqrt{3}( \mathcal{G}_z k_x +  \mathcal{G}_x k_z),  \\
\label{eq:Mxy_k1}
M^{\rm (I)}_{xy}(\bm{k})&=\sqrt{3}( \mathcal{G}_x k_y+  \mathcal{G}_y k_x).
\end{align}
In the case of the contraction between $\bm{k}$ and $\mathcal{Q}_{\mu\nu}$, the MQ is expressed as 
\begin{align}
\label{eq:Mu_k2}
M^{\rm (II)}_u(\bm{k})&= -3 \mathcal{Q}_{yz} k_x + 3\mathcal{Q}_{zx} k_y, \\
\label{eq:Mv_k2}
M^{\rm (II)}_v(\bm{k})&= \sqrt{3}(-\mathcal{Q}_{yz} k_x -\mathcal{Q}_{zx}k_y + 2 \mathcal{Q}_{xy}k_z), \\
\label{eq:Myz_k2}
M^{\rm (II)}_{yz}(\bm{k})&=\sqrt{3}\left[\sqrt{2}(\mathcal{Q}_{zz}-\mathcal{Q}_{yy}) k_x + \mathcal{Q}_{xy} k_y - \mathcal{Q}_{zx}k_z \right], \\
\label{eq:Mzx_k2}
M^{\rm (II)}_{zx}(\bm{k})&=\sqrt{3}\left[\sqrt{2}(\mathcal{Q}_{xx}-\mathcal{Q}_{zz}) k_y + \mathcal{Q}_{yz} k_z - \mathcal{Q}_{xy}k_x \right], \\
\label{eq:Mxy_k2}
M^{\rm (II)}_{xy}(\bm{k})&=\sqrt{3}\left[\sqrt{2}(\mathcal{Q}_{yy}-\mathcal{Q}_{xx}) k_z + \mathcal{Q}_{zx} k_x - \mathcal{Q}_{yz}k_y \right].  
\end{align}
As $\{M^{\rm (I)}_u(\bm{k}), M^{\rm (I)}_v(\bm{k}), M^{\rm (I)}_{yz}(\bm{k}), M^{\rm (I)}_{zx}(\bm{k}), M^{\rm (I)}_{xy}(\bm{k})\}$ and $\{M^{\rm (II)}_u(\bm{k}), M^{\rm (II)}_v(\bm{k}), M^{\rm (II)}_{yz}(\bm{k}), M^{\rm (II)}_{zx}(\bm{k}), M^{\rm (II)}_{xy}(\bm{k})\}$ are the same spatial property, their linear combination, e.g., $c_1 M^{\rm (I)}_u(\bm{k})+c_2 M^{\rm (II)}_u(\bm{k})$ where $c_1$ and $c_2$ are linear coefficients, are expected to appear once the MQ order occurs. 
 
The expressions in Eqs.~(\ref{eq:Mu_k1})-(\ref{eq:Mxy_k2}) indicate an emergent antisymmetric spin-orbital polarization with respect to $\bm{k}$ in the band structure. 
The product of $\bm{l}$ and $\bm{\sigma}$ in $\bm{\mathcal{G}}$ and $\mathcal{Q}_{\mu\nu}$ is higher-rank coupling to the atomic spin-orbit coupling $\bm{l}\cdot \bm{\sigma}$. 
The $\bm{k}$ dependence is qualitatively different with each other: 
The present coupling shows an antisymmetric $\bm{k}$ dependence, whereas the ordinary spin-orbit coupling does not. 
Besides, the coupling between the different components of $\bm{l}$ and $\bm{\sigma}$ can emerge in the MQ ordered state, e.g., $\mathcal{G}_x=l_y \sigma_z - l_z \sigma_y$ and $\mathcal{Q}_{xy} =(l_x \sigma_y+l_y \sigma_x)/\sqrt{2}$. 
Since $\mathcal{Q}_{\mu\nu}$ has the same symmetry property as the quadrupole, this spin-orbital polarization is regarded as the antisymmetric quadrupole splitting. 
Thus, the appearance of the MQ ordering connects between the momentum and the spin-orbital degree of freedom, which is the microscopic origin of the current-induced distortion, as will be discussed in Sec.~\ref{sec:Current-induced distortion}. 

The above antisymmetric spin-orbital polarization is similar to the antisymmetric spin polarization in a nonmagnetic noncentrosymmetric lattice system with the Rashba or Dresselhaus spin-orbit interaction. 
In the case of the nonmagnetic systems, the antisymmetric spin splittings appear in the form of $\bm{k} \times \bm{\sigma}$ for the polar crystal and $k_\mu \sigma_\nu$ for the chiral/gyrotropic crystal. 
Owing to the momentum dependence in the spin splitting, the spin orientation is locked at the particular direction at each wave vector $\bm{k}$, which is called the spin momentum locking~\cite{hsieh2009tunable}. 
In a similar way, the present antisymmetric spin-orbital polarization leads to the locking of the component of $l_{\mu}\sigma_\nu$ at the particular component at each $\bm{k}$. 
Thus, we term the antisymmetric spin-orbital polarization in the MQ ordered state as the spin-orbital momentum locking. 
It is noted that this spin-orbital momentum locking does not accompany the individual spin and orbital polarizations owing to the $\mathcal{PT}$ symmetry.
This situation can be regarded as the hidden spin polarization in the band structure, which is similar to that discussed in the staggered Rashba systems without local inversion symmetry at each lattice site~\cite{zhang2014hidden,Fu_PhysRevLett.115.026401,Razzoli_PhysRevLett.118.086402,gotlieb2018revealing,Huang_PhysRevB.102.085205,Ishizuka_PhysRevB.98.224510}, such as the zigzag~\cite{Yanase_JPSJ.83.014703,Hayami_doi:10.7566/JPSJ.84.064717,Sumita_PhysRevB.93.224507,cysne2021orbital}, honeycomb~\cite{Kane_PhysRevLett.95.226801,Hayami_PhysRevB.90.081115,Hayami_1742-6596-592-1-012131,hayami2016emergent,yanagi2017optical}, diamond~\cite{Fu_PhysRevLett.98.106803,Hayami_PhysRevB.97.024414,Ishitobi_doi:10.7566/JPSJ.88.063708}, and layered systems~\cite{Hayami_PhysRevB.90.024432,hitomi2014electric,hitomi2016electric,yao2017direct}. 
In contrast to the staggered Rashba systems, the present spin-orbital polarization with hidden spin and orbital polarizations is activated by the spontaneous MQ ordering irrespective of the specific lattice structure. 
It is noted that the hidden polarizations can be lifted easily by applying external magnetic field. 

Let us remark on the relevance with other multipole orderings. 
The antisymmetric spin-orbital momentum locking can also appear in the odd-parity magnetic and magnetic toroidal multipoles, such as the magnetic toroidal dipole. 
For instance, the contraction of $\bm{k}$ and $\bm{\mathcal{G}}$ includes the rank-1 polar tensor corresponding to the magnetic toroidal dipole, $\bm{k} \times \bm{\mathcal{G}}$.
There, however, is a clear difference in the band structure between the MQ and the magnetic toroidal dipole after tracing out the spin-orbital degree of freedom: The former shows the symmetric band structure, while the latter exhibits the antisymmetric one with respect to $\bm{k}$.

\section{Active magnetic-quadrupole system}
\label{sec:Active magnetic-quadrupole system}
In this section, we demonstrate that the MQ ordering gives rise to the spin-orbital momentum locking by analyzing the specific lattice models. 
We consider two intuitive systems including the MQ degree of freedom: 
One is the multi-orbital system where the MQ is activated by the spin-dependent hybridization between orbitals with different spatial parity in Sec.~\ref{sec:Multi-orbital system}. 
The other is the sublattice system where the MQ is activated by the antiferromagnetic ordering in Sec.~\ref{sec:Multi-sublattice system}.

\subsection{Multi-orbital system}
\label{sec:Multi-orbital system}

In this section, we consider the MQ ordering in the multi-orbital system. 
By constructing the multi-orbital model consisting of the $s$, $p_x$, and $p_y$ orbitals where the MQ degree of freedom is described by the spin-dependent $s$-$p$ hybridization in Sec.~\ref{sec:Model}, we discuss the electronic band structure under the MQ ordering in Sec.~\ref{sec:Band structure}. 
We show that the MQ ordering gives rise to a variety of the spin-orbital momentum locking depending on the type of MQ as introduced in Sec.~\ref{sec:Spin-orbital-momentum locking}. 
The spontaneous MQ ordering induces the effective spin-orbit interaction similar to the atomic relativistic spin-orbit coupling. 
In Sec.~\ref{sec:Spin splittings under magnetic field}, we show the effect of the magnetic field on the band structure in the MQ ordering. 
The symmetric and antisymmetric spin splittings in addition to the Zeeman splitting are induced by the magnetic field owing to the $\mathcal{PT}$ symmetry breaking. 
We discuss the relation between the spin-orbital momentum locking and physical responses by exemplifying two cross-correlated phenomena: current-induced distortion in Sec.~\ref{sec:Current-induced distortion} and the magnetoelectric effect in Sec.~\ref{sec:Magnetoelectric effect}

\subsubsection{Model}
\label{sec:Model}

\begin{figure}[t!]
\begin{center}
\includegraphics[width=1.0 \hsize]{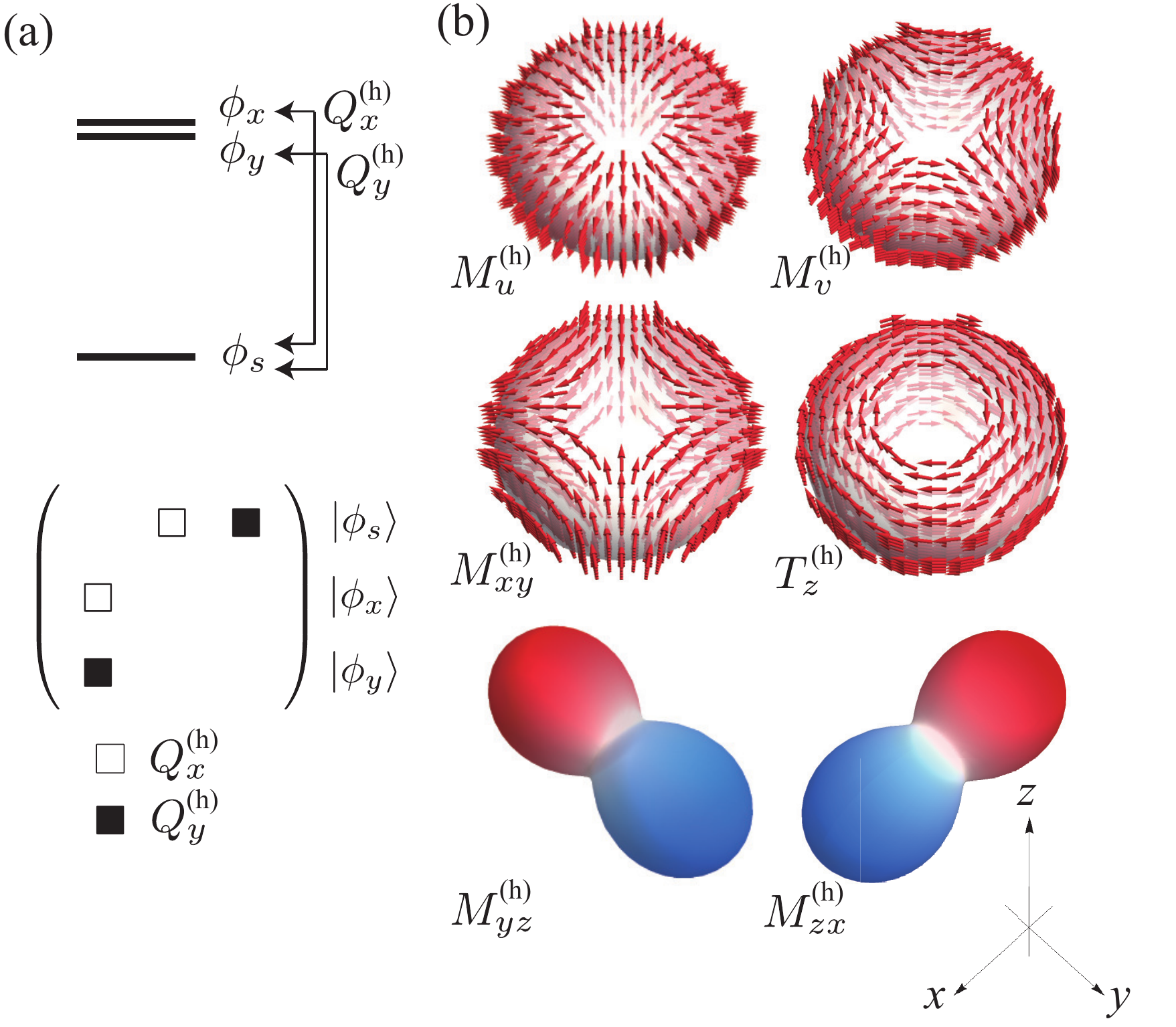} 
\caption{
\label{Fig:MultiOrbital}
(a) Multi-orbital model consisting of $s$, $p_x$, and $p_y$ orbitals whose wave functions are denoted as $\phi_s$, $\phi_x$, and $\phi_y$. 
The electric dipoles degree of freedom, $Q^{\rm (h)}_x$ and $Q^{\rm (h)}_y$, are defined by the off-diagonal elements between $\phi_s$ and $\phi_x$ (open squares) and $\phi_s$ and $\phi_y$ (closed squares).   
(b) The wave functions of the MQ $\{M^{\rm (h)}_u, M^{\rm (h)}_v, M^{\rm (h)}_{yz}, M^{\rm (h)}_{zx}, M^{\rm (h)}_{xy}\}$ and magnetic toroidal dipole $T^{\rm (h)}_z$ orderings in Eqs.~(\ref{eq:Mu_multiorbital})-(\ref{eq:Tz_multiorbital}). 
For visibility, we use the arrow for the angle distribution of the $xy$-spin moments for $M^{\rm (h)}_u$, $M^{\rm (h)}_v$, $M^{\rm (h)}_{xy}$, and $T^{\rm (h)}_z$, while the colormap is used for the $z$-spin moments for $M^{\rm (h)}_{yz}$ and $M^{\rm (h)}_{zx}$.
The shape represents the angle distributions of the electric charge density. 
}
\end{center}
\end{figure}

We construct a minimal multi-orbital model to describe the MQ degree of freedom. 
We consider a three-orbital model consisting of the $s$, $p_x$, and $p_y$ orbitals on a two-dimensional square lattice under the point group $D_{\rm 4h}$. 
We take the lattice constant as unity. 
The wave function of the $s$, $p_x$, and $p_y$ orbitals are represented by $\phi_s$, $\phi_{x}$, and $\phi_{y}$, respectively. 
Then, the tight-binding Hamiltonian for the basis $\{ \phi_{s\sigma}, \phi_{x\sigma}, \phi_{y\sigma} \}$ is given by 
\begin{align}
\label{eq:Ham_multiorbital}
\mathcal{H}=\sum_{\bm{k},\alpha,\beta,\sigma}c^{\dagger}_{\bm{k}\alpha \sigma} H^{\alpha\beta}_\sigma c_{\bm{k}\beta\sigma}, 
\end{align}
where $c^{\dagger}_{\bm{k}\alpha \sigma}$ ($c_{\bm{k}\alpha \sigma}$) is the creation (annihilation) operator of electrons at wave vector $\bm{k}$, orbital $\alpha=s, x, y$, and spin $\sigma$. 
The $3\times 3$ Hamiltonian matrix spanned by the basis $\{ \phi_{s\sigma}, \phi_{x\sigma}, \phi_{y\sigma} \}$ is given by 
\begin{align}
\label{eq:Ham_spxpy}
H_{\sigma}=\left(
\begin{array}{cccccc}
2t_s (c_x + c_y)& -2 i t_{sp} s_x& -2 i t_{sp} s_y\\
2 i t_{sp} s_x &  2 (t_{\sigma} c_x + t_{\pi} c_y) & 2 t_{xy} s_x s_y -i \sigma \lambda  \\
2 i t_{sp} s_y& 2 t_{xy} s_x s_y+ i \sigma  \lambda &  2 (t_{\sigma} c_y + t_{\pi} c_x) \\
\end{array}
\right),  
\end{align}
where $c_\eta= \cos k_{\eta}$ and $s_{\eta}=\sin k_{\eta}$ for $\eta=x, y$. 
There are five hopping parameters in the Hamiltonian in Eq.~(\ref{eq:Ham_spxpy}), $\{t_s, t_{\sigma}, t_{\pi}, t_{sp}, t_{xy}\}$ from the symmetry of the system; the nearest-neighbor hoppings between the $s$ orbitals $t_s$, $p$ orbitals $t_{\sigma}$ and $t_{\pi}$, and $s$-$p$ orbitals $t_{sp}$ and the next-nearest-neighbor hopping between the $p$ orbitals $t_{xy}$. 
$\lambda$ is the constant for the atomic spin-orbit coupling where the factor $1/2$ from the spin operator $\sigma/2$ is rescaled. 
We do not consider the atomic energy difference between $s$ and $p$ orbitals as well as the other hoppings for simplicity. 

The spinful $6\times 6$ Hamiltonian matrix spanned by $\{ \phi_{s\uparrow}, \phi_{x\uparrow}, \phi_{y\uparrow}, \phi_{s\downarrow}, \phi_{x\downarrow}, \phi_{y\downarrow} \}$ has 36 independent electronic degrees of freedom. 
In spinless space, there are 9 electronic degrees of freedom, whose irreducible representations are $2A^+_{1g} \oplus A^-_{2g} \oplus B^+_{1g} \oplus B^+_{2g} \oplus E^{\pm}_u$ where the superscript $\pm$ denotes the time-reversal parity. 
Among them, the irreducible representations $E^{+}_u$ and $E^{-}_u$ correspond to the odd-parity dipoles; the electric dipoles, $Q^{\rm (h)}_x$ and $Q^{\rm (h)}_y$, and magnetic toroidal dipoles, $T^{\rm (h)}_x$ and $T^{\rm (h)}_y$, respectively. 
Physically, they are expressed as the real and imaginary hybridizations between $s$ and $p$ orbitals, whose matrices are represented by
\begin{align}
Q^{\rm (h)}_x=
\left(
\begin{array}{ccc}
0 & 1 & 0 \\
1 & 0 & 0 \\
0 & 0 & 0 
\end{array}
\right), 
Q^{\rm (h)}_y=
\left(
\begin{array}{ccc}
0 & 0 & 1 \\
0 & 0 & 0 \\
1 & 0 & 0 
\end{array}
\right), \\
T^{\rm (h)}_x=
\left(
\begin{array}{ccc}
0 & i & 0 \\
-i & 0 & 0 \\
0 & 0 & 0 
\end{array}
\right), 
T^{\rm (h)}_y=
\left(
\begin{array}{ccc}
0 & 0 & i \\
0 & 0 & 0 \\
-i & 0 & 0 
\end{array}
\right),  
\end{align}
where the superscript (h) represents the hybrid multipole that are active in the hybridized orbitals~\cite{hayami2018microscopic}. 
See also Fig.~\ref{Fig:MultiOrbital}(a). 
It is noted that there is no MQ degree of freedom in spinless space. 

The MQ degree of freedom with $(\mathcal{P}, \mathcal{T})=(-1, -1)$ appears by considering the electronic degrees of freedom in spinful space, which is constructed from the product of the electric dipole with $(\mathcal{P}, \mathcal{T})=(-1, 1)$ and spin operator with $(\mathcal{P}, \mathcal{T})=(1, -1)$~\cite{hayami2018microscopic,kusunose2020complete}. 
Indeed, the product of the irreducible representation of the electric dipole $E^{+}_u$ and spin $A^-_{2g} \oplus E^{-}_g$ gives six odd-parity multipoles with time-reversal odd; $E^{+}_u\otimes (A^-_{2g} \oplus E^{-}_g) = A^{-}_{1u}\oplus A^{-}_{2u}\oplus B^{-}_{1u}\oplus B^{-}_{2u}\oplus E^-_{u}$. 
Among them, the five out of six multipole degrees of freedom are expressed as the MQs, whose expressions are given by 
\begin{align}
\label{eq:Mu_multiorbital}
A_{1u}^{-}:\quad&
M^{\rm (h)}_u = -Q^{\rm (h)}_x \sigma_x -Q^{\rm (h)}_y \sigma_y, \\
\label{eq:Mv_multiorbital}
B_{1u}^{-}:\quad&
M^{\rm (h)}_v = Q^{\rm (h)}_x \sigma_x - Q^{\rm (h)}_y \sigma_y, \\
\label{eq:Myz_multiorbital}
E_{u}^{-}:\quad&
M^{\rm (h)}_{yz} = Q^{\rm (h)}_y \sigma_z, \\
\label{eq:Mzx_multiorbital}
&
M^{\rm (h)}_{zx} = Q^{\rm (h)}_x \sigma_z, \\
\label{eq:Mxy_multiorbital}
B_{2u}^{-}:\quad&
M^{\rm (h)}_{xy} = Q^{\rm (h)}_x \sigma_y + Q^{\rm (h)}_y \sigma_x.
\end{align}
One can find that the expressions in Eqs.~(\ref{eq:Mu_multiorbital})-(\ref{eq:Mxy_multiorbital}) correspond to those in Eqs.~(\ref{eq:Mu})-(\ref{eq:Mxy}) in Sec.~\ref{sec:Expressions in real space} by replacing $(x, y, z)$ and $(m_x, m_y, m_z)$ with $(Q^{\rm (h)}_x, Q^{\rm (h)}_y, 0)$ and $(\sigma_x, \sigma_y, \sigma_z)$, respectively.
It is noted that $M^{\rm (h)}_{yz}$ and $M^{\rm (h)}_{zx}$ belong to the same irreducible representation as the inplane magnetic toroidal dipoles, $T_x$ and $T_y$, respectively, which will be discussed in details in Sec.~\ref{sec:Band structure}. 

The remaining one corresponds to the magnetic toroidal dipole degree of freedom $T^{\rm (h)}_{z}$, which belongs to the irreducible representation $A^{-}_{2u}$.  
The expression of $T^{\rm (h)}_{z}$ is given by the antisymmetric product between $\{Q^{\rm (h)}_x, Q^{\rm (h)}_y\}$ and $\{\sigma_x, \sigma_y\}$ as~\cite{hayami2018microscopic,yatsushiro2019atomic} 
\begin{align}
\label{eq:Tz_multiorbital}
A_{2u}^{-}:\quad T^{\rm (h)}_{z} &= Q^{\rm (h)}_x \sigma_y - Q^{\rm (h)}_y \sigma_x.  
\end{align}
The above six multipoles satisfy the orthogonal relation: ${\rm Tr}[X Y]=0$ where $X, Y=M^{\rm (h)}_u, M^{\rm (h)}_v, M^{\rm (h)}_{yz}, M^{\rm (h)}_{zx}, M^{\rm (h)}_{xy}, T^{\rm (h)}_{z}$ and $X \neq Y $. 
In the following, we consider the situation where one of the six multipoles are ordered by the electron correlation. 
The amplitude of the mean field is given by $h$. 
We show the schematic pictures of the wave functions with nonzero $M^{\rm (h)}_u$, $M^{\rm (h)}_v$, $M^{\rm (h)}_{yz}$, $M^{\rm (h)}_{zx}$, $M^{\rm (h)}_{xy}$, and $T^{\rm (h)}_{z}$ in Fig.~\ref{Fig:MultiOrbital}(b). 
The shape represents the electric charge density, while the arrows for $M^{\rm (h)}_u$, $M^{\rm (h)}_v$, $M^{\rm (h)}_{xy}$, and $T^{\rm (h)}_z$ [colors for $M^{\rm (h)}_{yz}$ and $M^{\rm (h)}_{zx}$] represent the angle distributions of the $xy$($z$)-spin moments, which well correspond to the schematic spin polarization in Fig.~\ref{Fig:MQ}.

\subsubsection{Band structure}
\label{sec:Band structure}

\begin{figure*}[htb!]
\begin{center}
\includegraphics[width=0.97 \hsize]{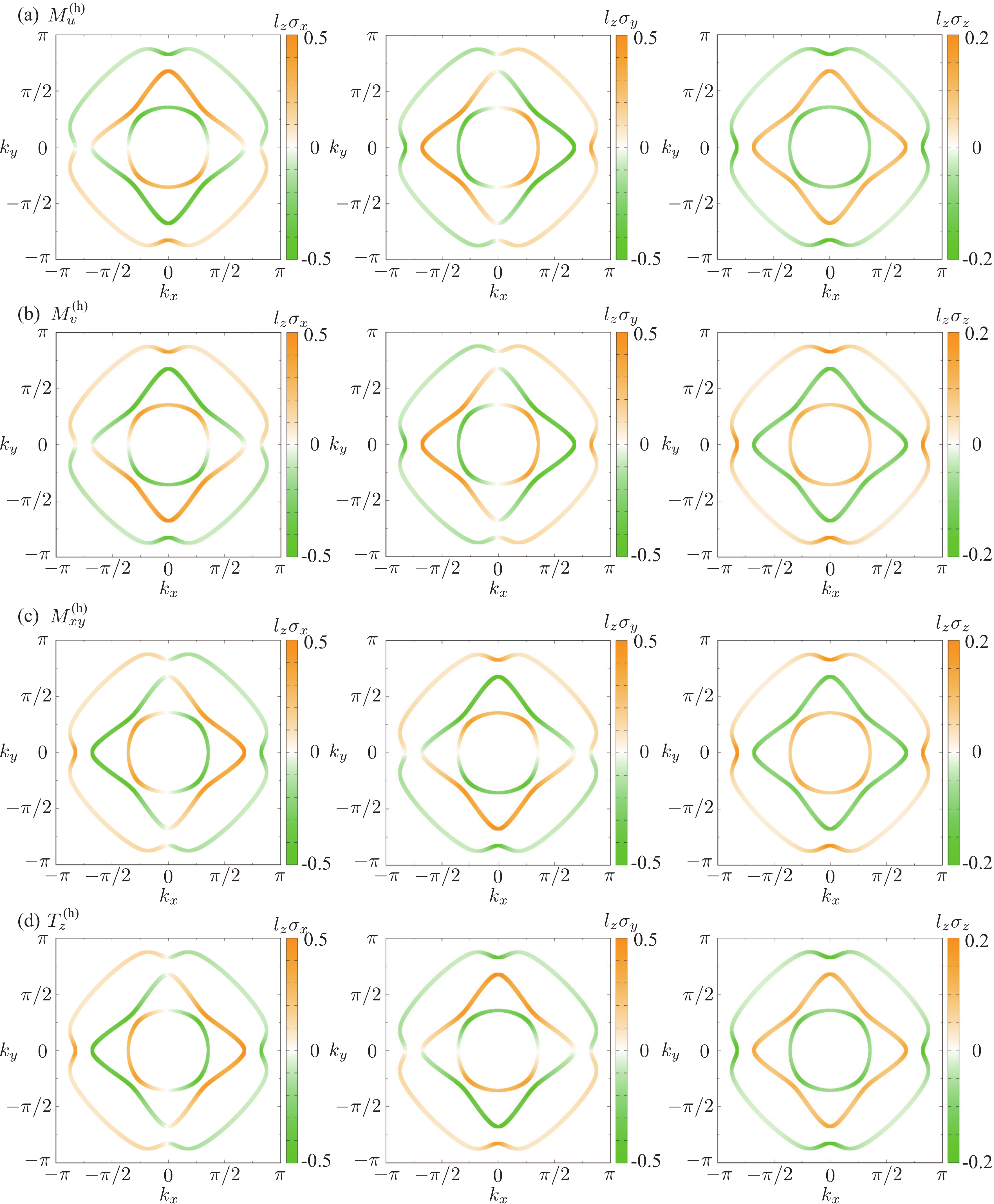} 
\caption{
\label{Fig:MultiOrbital_band}
The isoenergy surfaces at $\mu=1$, $t_{\sigma}=0.8$, $t_{\pi}=0.5$, $t_{xy}=0.3$, $t_s=1$, $t_{sp}=0.6$, and $\lambda=0$ in the (a) $M^{\rm (h)}_u$, (b) $M^{\rm (h)}_v$, (c) $M^{\rm (h)}_{xy}$, and (d) $T^{\rm (h)}_z$ states with the molecular field $h=0.3$. 
The colormap shows the spin-orbital polarization of the $l_z \sigma_x$ (left), $l_z \sigma_y$ (middle), and $l_z \sigma_z$ (right) components at each wave vector.   
}
\end{center}
\end{figure*}

\begin{figure}[htb!]
\begin{center}
\includegraphics[width=1.0 \hsize]{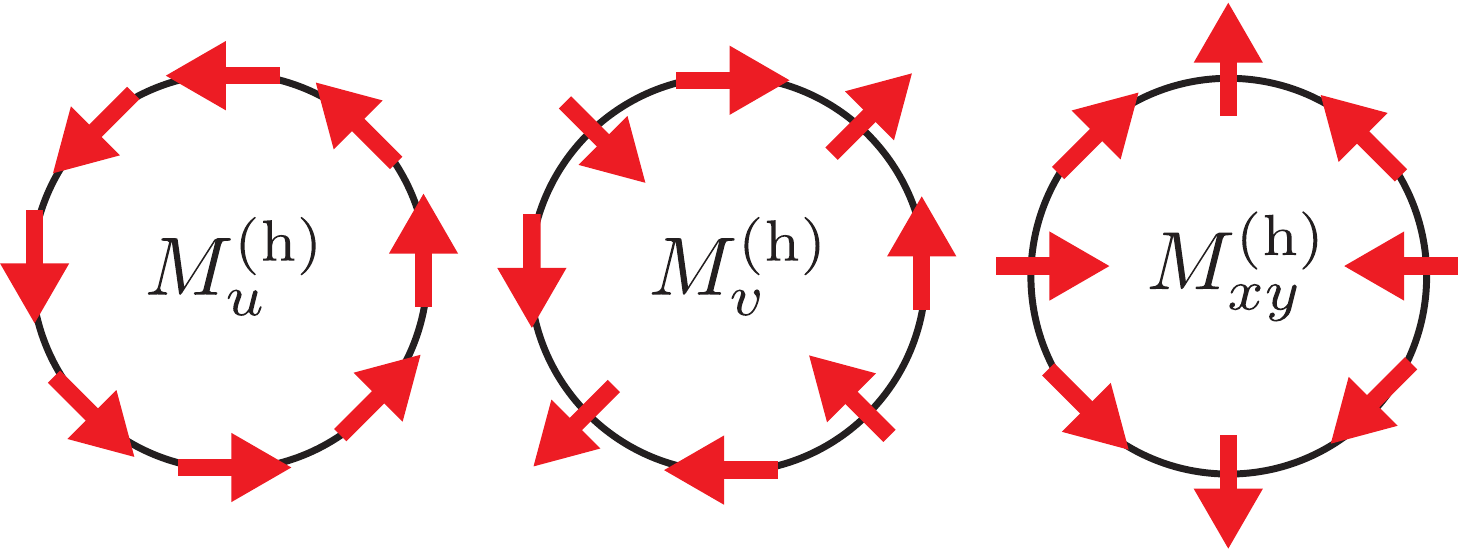} 
\caption{
\label{Fig:band_ponti}
The schematic spin-orbital momentum locking on the circular Fermi surfaces for the $M^{\rm (h)}_u$, $M^{\rm (h)}_v$, and $M^{\rm (h)}_{xy}$ ordered states. 
The arrow represents the direction of $(\mathcal{Q}_x, \mathcal{Q}_y)$. 
}
\end{center}
\end{figure}

\begin{figure}[htb!]
\begin{center}
\includegraphics[width=1.0 \hsize]{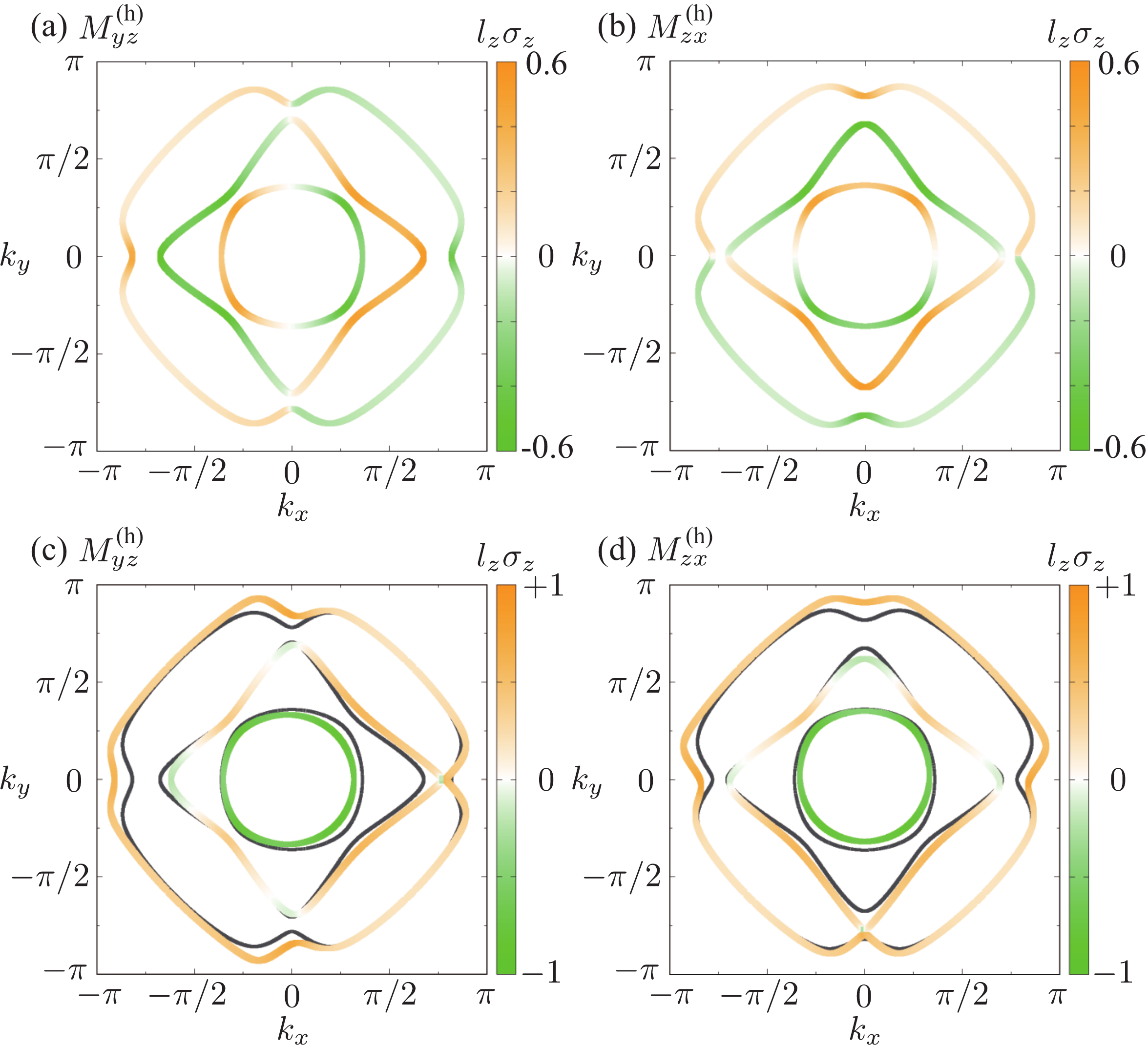} 
\caption{
\label{Fig:MultiOrbital_band_Myz}
The same plot with Fig.~\ref{Fig:MultiOrbital_band} in the (a,c) $M^{\rm (h)}_{yz}$ and (b,d) $M^{\rm (h)}_{zx}$ states. 
The same parameters are used in (a) and (b), and the effect of the spin-orbit coupling $\lambda$ is added as $\lambda=0.5$ in (c) and (d). 
The solid curves in (c) and (d) represent the isoenergy surfaces in (a) and (b), respectively. 
}
\end{center}
\end{figure}

We investigate the change of the electronic band structure in the presence of the MQ orderings in the multi-orbital model in Eq.~(\ref{eq:Ham_multiorbital}). 
As the system is two-dimensional ($k_z=0$) and the Hamiltonian has only the $z$ component of the angular momentum ($l_x=l_y=0$), the spin-orbital momentum locking in Eqs.~(\ref{eq:Mu_k1})-(\ref{eq:Mxy_k2}) reduces to 
\begin{align}
\label{eq:Muk_orbital}
M_u(\bm{k})&= -  l_z \sigma_y k_x +  l_z \sigma_x  k_y, \\
\label{eq:Mvk_orbital}
M_v(\bm{k})&= - l_z \sigma_y k_x -l_z \sigma_x k_y,  \\
\label{eq:Myzk_orbital}
M_{yz}(\bm{k})&=l_z \sigma_z  k_x,  \\
\label{eq:Mzxk_orbital}
M_{zx}(\bm{k})&=- l_z \sigma_z k_y,  \\
\label{eq:Mxyk_orbital}
M_{xy}(\bm{k})&= l_z \sigma_x  k_x -  l_z \sigma_y k_y, 
\end{align}
except for the numerical coefficient. 
Thus, the spin-orbital momentum locking with respect to the components of $l_z \sigma_x$, $l_z \sigma_y$, and $l_z \sigma_z$ are expected once the MQ order occurs.

Figures~\ref{Fig:MultiOrbital_band}(a)-\ref{Fig:MultiOrbital_band}(c) show the isoenergy surfaces in the band structure at the chemical potential $\mu=1$ for the $M^{(\rm h)}_u$, $M^{(\rm h)}_v$, and $M^{(\rm h)}_{xy}$ ordered states, respectively. 
The model parameters are taken as $t_{\sigma}=0.8$, $t_{\pi}=0.5$, $t_{xy}=0.3$, $t_s=1$, $t_{sp}=0.6$, and $h=0.3$. 
We neglect the effect of the atomic relativistic spin-orbit coupling $\lambda=0$ unless otherwise stated.
The spin-orbital polarizations $l_z \sigma_x$, $l_z \sigma_y$, and $l_z \sigma_z$, are calculated at each $\bm{k}$, which are shown in the left, middle, and right columns in Fig.~\ref{Fig:MultiOrbital_band}, respectively.  
Each band is doubly degenerate owing to the $\mathcal{PT}$ symmetry. 

The results clearly indicate the emergence of the spin-orbital momentum locking expected from the symmetry argument in Eqs~(\ref{eq:Muk_orbital})-(\ref{eq:Mxyk_orbital}) in each ordered state.
The antisymmetric spin-orbital polarizations of $l_z \sigma_x$ and $l_z \sigma_y$ occur along the $k_y$ and $k_x$ directions, respectively, in the case of the $M^{(\rm h)}_u$ ordered state in Fig.~\ref{Fig:MultiOrbital_band}(a). 
Similarly, the antisymmetric spin-orbital polarizations in Eqs.~(\ref{eq:Mvk_orbital}) and (\ref{eq:Mxyk_orbital}) are found in the $M^{(\rm h)}_v$ and $M^{(\rm h)}_{xy}$ ordered states, as shown in Figs.~\ref{Fig:MultiOrbital_band}(b) and \ref{Fig:MultiOrbital_band}(c), respectively. 

The important hopping parameters for the spin-orbital momentum locking are easily extracted by evaluating the following quantity at each wave vector $\bm{k}$, $\mathcal{O}_{\mu}(\bm{k}) ={\rm Tr}[e^{-\beta \mathcal{H}_{\bm{k}}} l_z \sigma_\mu]$ for $\mu=x,y,z$ and $\mathcal{H}=\sum_{\bm{k}}\mathcal{H}_{\bm{k}}$, where $\beta$ is the inverse temperature. 
In the high-temperature expansion of $\mathcal{O}_{\mu}(\bm{k})$, the necessary hopping parameters for the spin-orbital momentum locking are systematically obtained~\cite{Hayami_PhysRevB.101.220403,Hayami_PhysRevB.102.144441}. 
For the $M^{(\rm h)}_u$ ordered state, the lowest-order contributions of $\mathcal{O}_{x}(\bm{k})$ and $\mathcal{O}_{y}(\bm{k})$ are given by $-h t_{sp} \sin k_y$ and $h t_{sp} \sin k_x$, respectively, which indicates that the antisymmetric spin-orbital polarization is induced by the effective coupling between the order parameter $h$ and the $s$-$p$ hopping $t_{sp}$. 

Notably, there is a symmetric spin-orbital polarization in terms of the $l_z \sigma_z$ component even without the atomic spin-orbit coupling, as shown in Figs.~\ref{Fig:MultiOrbital_band}(a)-\ref{Fig:MultiOrbital_band}(c). 
This is because the order parameters in Eqs.~(\ref{eq:Mu_multiorbital}), (\ref{eq:Mv_multiorbital}), and (\ref{eq:Mxy_multiorbital}) are described by two spin components, $\sigma_x$ and $\sigma_y$. 
In this case, the term proportional to $l_z \sigma_z$ appears as the even-order product of the mean-field term in the expansion of $\mathcal{O}_{z}(\bm{k})$. 
Indeed, the lowest-order contribution of $\mathcal{O}_{z}(\bm{k})$ is proportional to $h^2$. 
The opposite sign of $l_z \sigma_z$ between the $M^{\rm (h)}_u$ and the other two ordered states is due to the opposite vorticity of the vector $\bm{\mathcal{Q}}(\bm{k})=(\langle l_z \sigma_x (\bm{k}) \rangle, \langle  l_z \sigma_y (\bm{k}) \rangle, \langle l_z \sigma_z (\bm{k}) \rangle)$ in $\bm{k}$ space: The direction of $(\mathcal{Q}_x, \mathcal{Q}_y)$ shows a (counter)clockwise rotation for the $M^{\rm (h)}_v$ and $M^{\rm (h)}_{xy}$ ($M^{\rm (h)}_u$) ordered states for the counterclockwise path on the circular Fermi surfaces, as schematically shown in Fig.~\ref{Fig:band_ponti}
As the almost uniform distribution of $l_z \sigma_z$ in $\bm{k}$ space resembles the atomic spin-orbit coupling, this is regarded as the emergent spin-orbit coupling arising from the MQ ordering. 
Thus, the magnitude of the atomic spin-orbit coupling can be controlled by the magnitude of the MQ order parameter.

Similar to the MQ ordering, the magnetic toroidal dipole ordering $T^{\rm (h)}_z$ in Eq.~(\ref{eq:Tz_multiorbital}) also shows the antisymmetric spin-orbital polarization, as shown in Fig.~\ref{Fig:MultiOrbital_band}(d). 
The functional form of the antisymmetric spin-orbital polarization is represented by $l_z \sigma_x  k_x +  l_z \sigma_y k_y $, which is obtained by replacing $\sigma_x$ with $-\sigma_x$ in the expression of $M_{xy}(\bm{k})$ in Eq~(\ref{eq:Mxyk_orbital}). 
Although the asymmetric bottom shift along the $k_z$ direction is expected with the onset of $T^{\rm (h)}_z$, it does not appear in the present two-dimensional system. 

The remaining MQs, $M^{\rm (h)}_{yz}$ and $M^{\rm (h)}_{zx}$, also show the spin-orbital momentum locking. 
The antisymmetric spin-orbital polarization of $l_z \sigma_z$ appears with respect to the $k_x$ ($k_y$) direction in the $M^{\rm (h)}_{yz}$ ($M^{\rm (h)}_{zx}$) ordered state, as shown in Fig.~\ref{Fig:MultiOrbital_band_Myz}(a) [Fig.~\ref{Fig:MultiOrbital_band_Myz}(b)]. 
In contrast to the result in Fig.~\ref{Fig:MultiOrbital_band}, there is no symmetric polarization of $l_z \sigma_z$ owing to the one spin component in Eqs.~(\ref{eq:Myz_multiorbital}) and (\ref{eq:Mzx_multiorbital}).   

It is noteworthy that $M^{\rm (h)}_{yz}$ and $M^{\rm (h)}_{zx}$ belong to the same irreducible representations of $T^{\rm (h)}_x$ and $T^{\rm (h)}_y$ under the point group $D_{4 {\rm h}}$. 
Nevertheless, there is no antisymmetric band bottom shift along the $k_x$ and $k_y$ directions in Figs.~\ref{Fig:MultiOrbital_band_Myz}(a) and \ref{Fig:MultiOrbital_band_Myz}(b). 
The antisymmetric band deformation appears only when introducing the atomic spin-orbit coupling $\lambda$ in Eq.~(\ref{eq:Ham_spxpy}), as shown in Figs.~\ref{Fig:MultiOrbital_band_Myz}(c) and \ref{Fig:MultiOrbital_band_Myz}(d). 
Indeed, the effective coupling between $\lambda$ and $h t_{sp}$ appears in the expansion of ${\rm Tr}[e^{-\beta \mathcal{H}_{\bm{k}}}]$; $-h t_{sp} \lambda \sin k_x $ for the $M^{\rm (h)}_{yz}$ state and $h t_{sp} \lambda \sin k_y $ for the $M^{\rm (h)}_{zx}$ state. 
This result indicates the magnetic toroidal dipole, $T^{\rm (h)}_x$ and $T^{\rm (h)}_y$, are secondary induced in the presence of the spin-orbit coupling in the MQ state.

\subsubsection{Spin splittings under magnetic field}
\label{sec:Spin splittings under magnetic field}

\begin{figure*}[htb!]
\begin{center}
\includegraphics[width=1.0 \hsize]{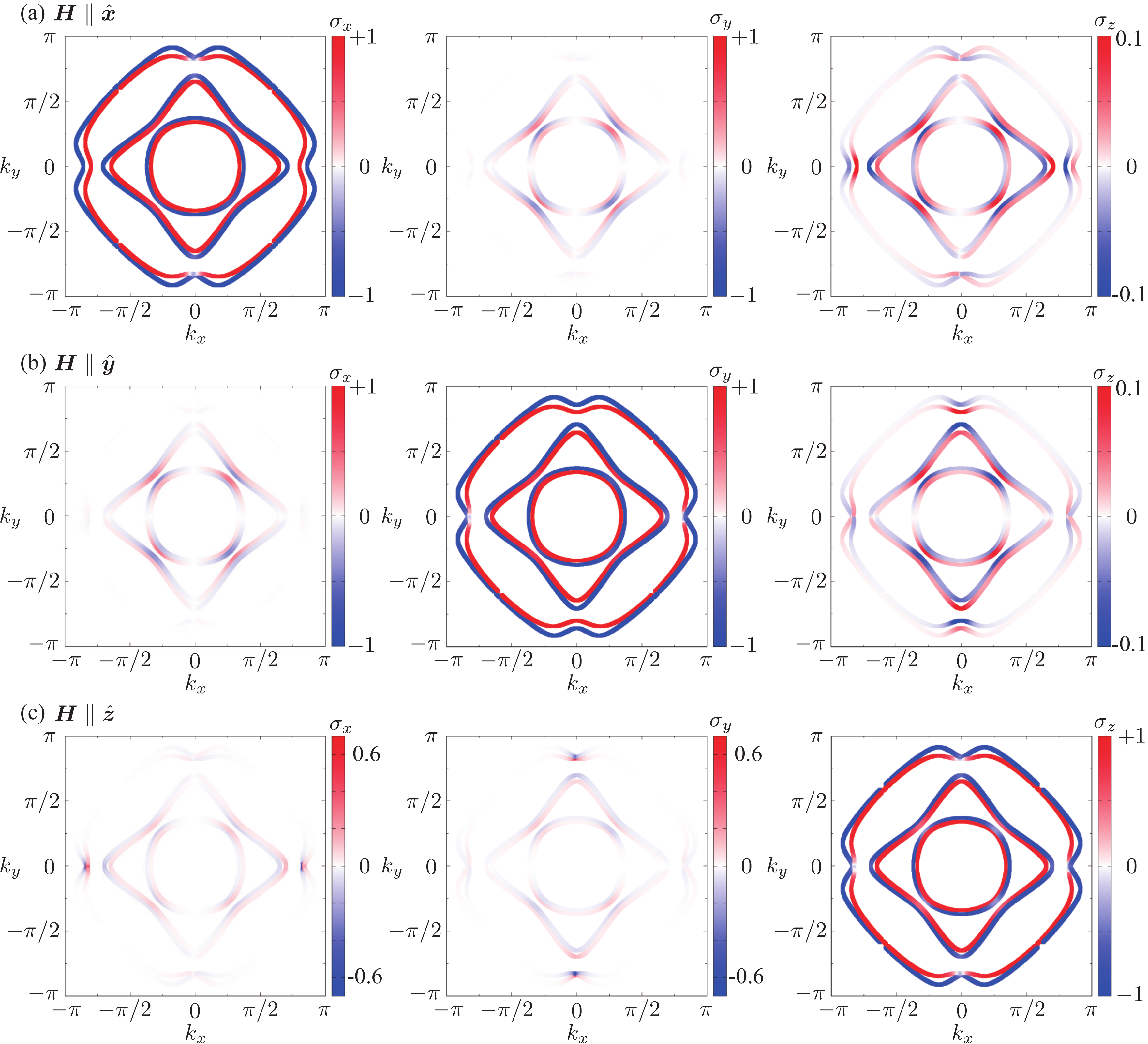} 
\caption{
\label{Fig:MultiOrbital_band_H}
The isoenergy surfaces in the $M^{\rm (h)}_{xy}$ state in the external magnetic field along the (a) $x$, (b) $y$, and (c) $z$ directions with $|\bm{H}|=0.1$
The colormap shows the spin polarization of the $x$ (left), $y$ (middle), and $z$ (right) components at each wave vector.   
The other parameters are the same as Fig.~\ref{Fig:MultiOrbital_band}. 
}
\end{center}
\end{figure*}

Although the MQ ordered state exhibits the antisymmetric spin-orbital polarization, it does not show any spin splittings owing to the presence of the $\mathcal{PT}$ symmetry. 
The degeneracy is lifted by the $\mathcal{PT}$-breaking field, which results in additional momentum-dependent spin splittings in the band structure. 
Hereafter, we demonstrate that various symmetric and antisymmetric spin splittings are induced by an external magnetic field by focusing on the $M^{\rm (h)}_{xy}$ ordered state. 
We introduce the Zeeman coupling to spin, $-\bm{H}\cdot \sum_{i \alpha \sigma \sigma'} c^{\dagger}_{i\alpha \sigma} \bm{\sigma}_{\sigma\sigma'}c_{i\alpha \sigma'}$, where we neglect the Zeeman coupling to orbital angular momentum without loss of generality. 

Figures~\ref{Fig:MultiOrbital_band_H}(a)-\ref{Fig:MultiOrbital_band_H}(c) show the isoenergy surfaces in the presence of the magnetic field along the (a) $x$, (b), $y$, and (c) $z$ directions with the magnitude of $|\bm{H}|\equiv H =0.1$. 
The three panels in each figure correspond to the spin polarization of $\sigma_x$, $\sigma_{y}$, and $\sigma_z$. 
The other parameters are the same as those in the previous section. 

When the magnetic field is turned on along the $x$ direction, the spin polarization emerges owing to the $\mathcal{PT}$ symmetry breaking, as shown in Fig.~\ref{Fig:MultiOrbital_band_H}(a), although the $\bm{k}$ dependence of the spin polarization is different for the different spin components. 
The $x$-spin component parallel to the magnetic field shows the ordinary Zeeman splitting, whereas the $y$ and $z$ spin components perpendicular to the magnetic field give rise to the symmetric and antisymmetric spin splittings, whose functional forms are given by $k_x k_y \sigma_y$ and $k_x \sigma_z$ in the limit of $\bm{k} \to \bm{0}$, respectively. 

Reflecting the different form of spin splittings, the necessary model parameters are different. 
We extract the essential model parameters for the spin splittings by calculating $\mathcal{S}_\mu(\bm{k})={\rm Tr}[e^{-\beta \mathcal{H}_{\bm{k}}}  \sigma_\mu]$~\cite{Hayami_PhysRevB.101.220403,Hayami_PhysRevB.102.144441}. 
In the $y$-spin component, the lowest-order contribution in $\mathcal{S}_y(\bm{k})$ for $\beta \mathcal{H}_{\bm{k}} \ll 1$ is given by $h^2 H t_{xy}   \sin k_x \sin k_y$ or $h^2 H t^2_{sp}   \sin k_x \sin k_y$, whereas that in the $z$-spin component $\mathcal{S}_z(\bm{k})$ is given by $h^3 H  t_{sp} \sin k_x $. 
From these expressions, one can find that the $s$-$p$ hopping $t_{sp}$ is necessary for the antisymmetric spin splitting, while it is not for the symmetric spin splitting.
Moreover, the domain formation is irrelevant (relevant) to the (anti)symmetric spin splitting as $\mathcal{S}_y(\bm{k})$ [$\mathcal{S}_z(\bm{k})$] is proportional to $h^2$ ($h^3$). 
These additional $\bm{k}$-dependent spin splittings are related to the active multipoles: The symmetric spin splitting like $k_x k_y \sigma_y$ corresponds to the magnetic toroidal quadrupole with the $zx$ component and the antisymmetric spin splitting like $k_x \sigma_z$ corresponds to the electric dipole along the $y$ direction~\cite{Hayami_PhysRevB.98.165110}. 
In particular, the latter electric dipole induced by the magnetic field implies the magnetoelectric effect in metals, which is relevant to the discussion in Sec.~\ref{sec:Magnetoelectric effect}.  

The similar band modulations also occur under the magnetic field along the $y$ and $z$ directions, as shown in Figs.~\ref{Fig:MultiOrbital_band_H}(b) and \ref{Fig:MultiOrbital_band_H}(c), respectively. 
For $\bm{H} \parallel \hat{\bm{y}}$, the symmetric (antisymmetric) spin splitting in the form of $k_x k_y$ ($k_y$) is found in the $x$($z$)-spin component in addition to the Zeeman splitting in the $y$-spin component. 
This indicates that the $yz$ component of the magnetic toroidal quadrupole and the $x$ component of the electric dipole are activated by the magnetic field along the $y$ direction. 

For $\bm{H} \parallel \hat{\bm{z}}$, the antisymmetric spin splitting occurs in both $\sigma_x$ and $\sigma_y$ components in Fig.~\ref{Fig:MultiOrbital_band_H}(c), whose functional form is represented by $k_x \sigma_x - k_y \sigma_y$. 
Indeed, by calculating $\mathcal{S}_x$ and $\mathcal{S}_y$, we obtain the coupling form as $h^3 H t_{sp} (\sin k_x \sigma_x - \sin k_y \sigma_y)$. 
It is noteworthy that this type of antisymmetric spin splitting indicates the active axial electric toroidal quadrupole  rather than the polar electric dipole. 
The appearance of the electric toroidal quadrupole is related with the optical rotation.

\subsubsection{Current-induced distortion}
\label{sec:Current-induced distortion}

Next, we discuss physical phenomena related by the spin-orbital momentum locking under the MQ ordering.
We here consider the piezo-electric effect where the symmetric distortion $\epsilon_{\zeta}$ is induced by the electric field $E_{\nu}$, i.e., $\epsilon_{\zeta} = \sum_\nu \Lambda_{\zeta \nu} E_\nu$ for $\zeta=u,v,yz,zx,xy$. 
The current-induced distortion tensor $\Lambda_{\zeta\nu}$ is calculated by the linear response theory as~\cite{hayami2016emergent,Watanabe_PhysRevB.96.064432,yatsushiro2020odd} 
\begin{align}
 \label{eq:DE}
\Lambda_{\zeta\nu} & = 
 \sum_{\bm k} \sum_{pq} \Pi_{pq}({\bm k}) Q_{\zeta{\bm k}}^{pq}v_{\nu{\bm k}}^{qp},
\end{align}
where 
\begin{align}
\Pi_{pq}({\bm k}) =\frac{e\hbar}{iV} \frac{f[\varepsilon_p({\bm k})] - f[\varepsilon_q({\bm k})]}{[\varepsilon_p({\bm k})-\varepsilon_q({\bm k})][\varepsilon_p({\bm k})-\varepsilon_q({\bm k})+i\hbar \delta]},
\end{align}
with the eigenenergy $\varepsilon_p({\bm k})$ and the Fermi distribution function $f[\varepsilon_p({\bm k})]$. 
$e$ is the electron charge, $\hbar = h/2\pi$ is the Plank constant divided by $2\pi$, $V$ is the system volume, and $\delta$ is the broadening factor. 
${Q}^{pq}_{\zeta {\bm k}}= \braket{p{\bm k}|\tilde{Q}_\zeta|q{\bm k}}$ and $v_{\nu{\bm k} }^{pq}=\braket{p{\bm k}|v_{\nu{\bm k}}|q{\bm k}}$ are the matrix elements of electric quadrupole and velocity $\hat{v}_{\mu{\bm k}} = \partial \hat{\mathcal{H}}/(\hbar \partial k_\mu)$, respectively. 
We regard $\l_\mu \sigma_\nu$ as the electric quadrupole degree of freedom $\tilde{Q}_\zeta$ from the symmetry viewpoint. 
We take $e=\hbar=1$, $\delta=0.01$, and the temperature $T=1/\beta=0.01$ in the following calculations.

The current-induced distortion tensor $\Lambda_{\zeta\nu}$ becomes nonzero in the absence of the spatial inversion symmetry in the system. 
$\Lambda_{\zeta\nu}$ consists of two parts: One is the intraband Fermi surface contribution $p=q$ and the other is the interband Fermi sea contribution $p \neq q$, where their time-reversal parity is opposite with each other. 
Reflecting the different time-reversal properties, the relevant multipoles are different in each contribution. 
The odd-parity magnetic and magnetic toroidal multipoles contribute to the intraband process, while the odd-parity electric and electric toroidal multipoles contribute to the interband process.  
In the present model in Eq.~(\ref{eq:Ham_spxpy}), the intraorbital contribution plays an important role in $\Lambda_{\zeta\nu}$, as the odd-parity electric and electric toroidal multipoles are not activated in the MQ ordered state owing to the $\mathcal{PT}$ symmetry. 

\begin{table}[t!]
\caption{
The irreducible representation (irrep.) of the magnetic and magnetic toroidal multipoles from the rank 0 to 3 under the point group $D_{\rm 4h}$. 
The nonzero matrix elements of the current-induced distortion tensor $\Lambda$ and the magnetoelectric tensor $\alpha$ are also shown. 
}
\label{tab_mp}
\centering
\begin{tabular}{lcccc}
\hline\hline
irrep. &  multipoles & $\Lambda_{\eta \mu}$ & $\alpha_{\mu \nu}$  \\
\hline 
$A^-_{1u}$ & $M_0$, $M_u$ & $\Lambda_{yzx}=-\Lambda_{zxy}$ & $\alpha_{xx}=\alpha_{yy}, \alpha_{zz}$ \\
$A^-_{2u}$ & $T_z$, $T^\alpha_z$ & $\Lambda_{zxx}=\Lambda_{yzy}$, $\Lambda_{uz}$ & $\alpha_{xy}=-\alpha_{yx}$ \\
$B^-_{1u}$ & $M_v$, $T_{xyz}$  & $\Lambda_{yzx}=\Lambda_{zxy}$, $\Lambda_{xyz}$ & $\alpha_{xx}=-\alpha_{yy}$ \\
$B^-_{2u}$ & $M_{xy}$,  $T_z^\beta$ & $\Lambda_{zxx}=-\Lambda_{yzy}$, $\Lambda_{vz}$ & $\alpha_{xy}=\alpha_{yx}$ \\
$E^-_{u}$ & $M_{yz}$, $T_x$, $T_x^\alpha$, $T_x^\beta$ & $\Lambda_{ux}$, $\Lambda_{vx}$, $\Lambda_{xyy}$, $\Lambda_{zxz}$ & $\alpha_{yz}$, $\alpha_{zy}$ \\
& $M_{zx}$, $T_y$, $T_y^\alpha$, $T_y^\beta$ & $\Lambda_{uy}$, $\Lambda_{vy}$, $\Lambda_{xyx}$, $\Lambda_{yzz}$ & $\alpha_{zx}$, $\alpha_{xz}$ \\
\hline \hline
\end{tabular}
\end{table}

\begin{figure}[htb!]
\begin{center}
\includegraphics[width=0.82 \hsize]{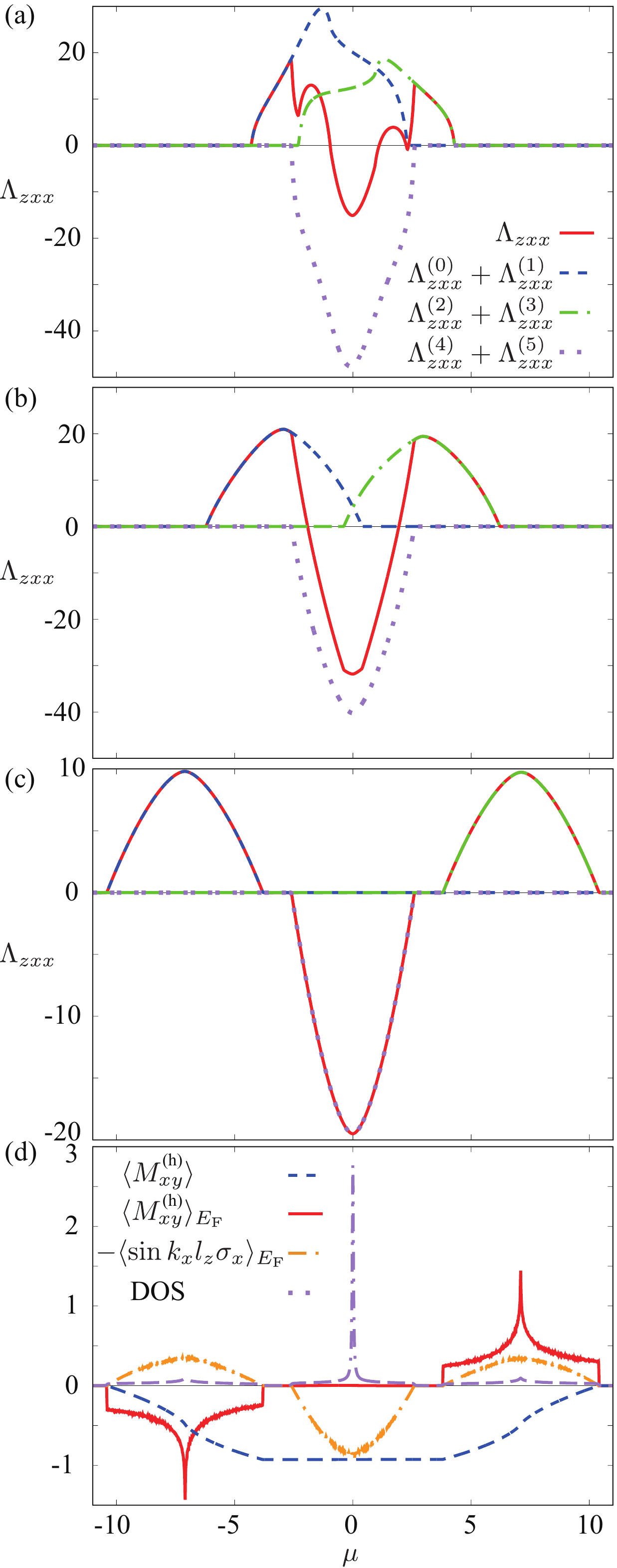} 
\caption{
\label{Fig:MultiOrbital_distortion}
(a)-(c) $\mu$ dependence of the coefficient of the current-distortion correlation, $\Lambda_{zxx}$, in the $M^{\rm (h)}_{xy}$ state at temperature $T=0.01$ and the damping factor $\delta=0.01$ for (a) $h=0.5$, (b) $2$, and (c) $5$. 
$\Lambda^{(p)}_{zxx}$ represents the intraband contribution from the $p$th band. 
The other parameters are the same as Fig.~\ref{Fig:MultiOrbital_band}. 
(d) $\mu$ dependences of the order parameter $\langle M^{\rm (h)}_{xy} \rangle$, its contribution near the Fermi level $\langle M^{\rm (h)}_{xy} \rangle_{E_{\rm F}}$, the spin-orbital polarization near the Fermi level $\langle \sin k_x l_z \sigma_x \rangle_{E_{\rm F}}$, and the density of states (DOS). 
The arbitrary unit is used for $\langle M^{\rm (h)}_{xy} \rangle_{E_{\rm F}}$ and $\langle \sin k_x l_z \sigma_x \rangle_{E_{\rm F}}$. 
}
\end{center}
\end{figure}

\begin{figure}[htb!]
\begin{center}
\includegraphics[width=1.0 \hsize]{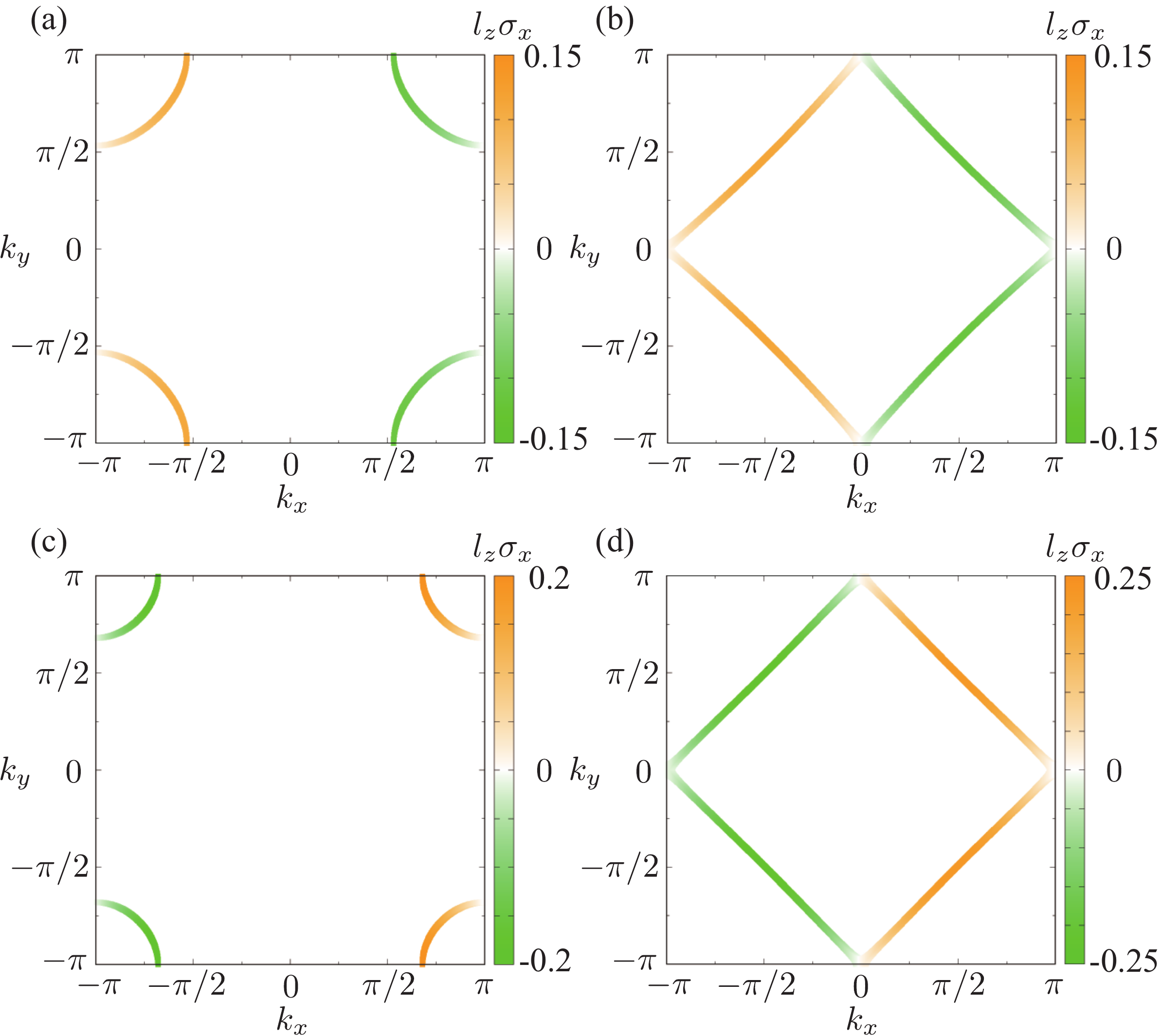} 
\caption{
\label{Fig:MultiOrbital_distortion_FS}
The $l_z \sigma_x$ polarization of the isoenergy surfaces in the $M_{xy}$ state for (a) $\mu=-9$, (b) $\mu=-7.1$, (c) $\mu=-2$, and (d) $\mu=0$.
The other parameters are $t_{\sigma}=0.8$, $t_{\pi}=0.5$, $t_{xy}=0.3$, $t_s=1$, $t_{sp}=0.6$, and $h=5$. 
}
\end{center}
\end{figure}

The 15 independent matrix elements of $\Lambda_{\zeta\nu}$ are characterized by the active rank 1-3 odd-parity multipoles with time-reversal odd: the magnetic toroidal dipole $\{T_x, T_y, T_z\}$, MQ $\{M_u, M_v, M_{yz}, M_{zx}, M_{xy} \}$, and magnetic toroidal octupole $\{ T_{xyz}, T^\alpha_x, T^\alpha_y, T^\alpha_z, T^\beta_x, T^\beta_y, T^\beta_z\}$. 
By using the multipole notation, the matrix form of $\Lambda_{\zeta\nu}$ is represented by~\cite{Hayami_PhysRevB.98.165110} 
\begin{widetext}
\begin{align}
&
\Lambda=
\begin{pmatrix}
T_{x}+M_{yz}-T_{x}^{\alpha}-T_{x}^{\beta} & T_{y}-M_{zx}-T_{y}^{\alpha}+T_{y}^{\beta} & -2T_{z}+2T_{z}^{\alpha} \\
-3T_{x}+M_{yz}+3T_{x}^{\alpha}-T_{x}^{\beta} & 3T_{y}+M_{zx}-3T_{y}^{\alpha}-T_{y}^{\beta} & -2M_{xy}+2T_{z}^{\beta} \\
-3M_{u}-M_{v}+T_{xyz} & -3T_{z}-M_{xy}-2T_{z}^{\alpha}-2T_{z}^{\beta} & -3T_{y}+M_{zx}-2T_{y}^{\alpha}+2T_{y}^{\beta} \\
-3T_{z}+M_{xy}-2T_{z}^{\alpha}+2T_{z}^{\beta} & 3M_{u}-M_{v}+T_{xyz} & -3T_{x}-M_{yz}-2T_{x}^{\alpha}-2T_{x}^{\beta} \\
-3T_{y}-M_{zx}-2T_{y}^{\alpha}-2T_{y}^{\beta} & -3T_{x}+M_{yz}-2T_{x}^{\alpha}+2T_{x}^{\beta} & 2M_{v}+T_{xyz}
\end{pmatrix},
\end{align}
\end{widetext}
where the row (column) of the matrix represents the component of $\{E_x, E_y, E_z\}$ ($\{\epsilon_{u}, \epsilon_{v}, \epsilon_{yz}, \epsilon_{zx}, \epsilon_{xy}\}$). 

Once the MQ ordering occurs, nonzero $\Lambda_{\zeta\mu}$ is obtained according to the types of the orderings. 
It is noted that the magnetic toroidal dipole and/or octupole belonging to the same irreducible representation of the MQ can be additionally activated, which also contributes to nonzero $\Lambda_{\zeta\mu}$. 
For example, in the case of $M_{xy}$ ordering, the magnetic toroidal octupole $T_z^\beta$ is simultaneously activated, which indicates two independent matrix elements in $\Lambda_{\zeta\mu}$, $\Lambda_{zxx}=-\Lambda_{yzy}$ and $\Lambda_{vz}$. 
The relation between the MQ ordering and nonzero $\Lambda$ in each irreducible representation is listed in Table~\ref{tab_mp}. 
In the following, we discuss the behavior of $\Lambda$ by focusing on the $M^{\rm (h)}_{xy}$ ordered state in the model in Eq.~(\ref{eq:Ham_multiorbital}).

Figure~\ref{Fig:MultiOrbital_distortion} shows the $\mu$ dependence of $\Lambda_{zxx}(=-\Lambda_{yzy})$ (solid red lines) in the $M^{\rm (h)}_{xy}$ ordered state. 
The results for different $h$ are plotted in Fig.~\ref{Fig:MultiOrbital_distortion}(a) at $h=0.5$, Fig.~\ref{Fig:MultiOrbital_distortion}(b) at $h=2$, and Fig.~\ref{Fig:MultiOrbital_distortion}(c) at $h=5$. 
The hopping parameters are $t_{\sigma}=0.8$, $t_{\pi}=0.5$, $t_{xy}=0.3$, $t_s=1$, $t_{sp}=0.6$, and $\lambda=0$, which are the same as those in Sec.~\ref{sec:Band structure}. 
It is noted that $\Lambda_{vz}=0$ because of the two-dimensional system. 

$\Lambda_{zxx}$ takes a finite value for nonzero $h$, as shown in Figs.~\ref{Fig:MultiOrbital_distortion}(a)-\ref{Fig:MultiOrbital_distortion}(c). 
However, it vanishes in the insulating region without the Fermi surfaces, e.g., $2.7<|\mu|<3.7$ for $h=5$ in Fig.~\ref{Fig:MultiOrbital_distortion}(c), since the intraband process at the Fermi surface is important in the presence of the MQ, as mentioned above. 
The overall behavior of $\Lambda_{zxx}$ against $\mu$ is similar: 
$\Lambda_{zxx}$ shows a positive value for the small Fermi surface (small electron/hole filling), while it becomes negative for the large Fermi surface (close to half filling). 
For larger $h=2$ and $5$, $\Lambda_{zxx}$ is characterized by two broad maxima and one broad minimum, whose positions become closer to the eigenvalues of the mean-field Hamiltonian, i.e., $\pm \sqrt{2}h, 0$, for larger $h$. 
Meanwhile, there are four maxima and three minima for $h=0.5$ in Fig.~\ref{Fig:MultiOrbital_distortion}(a). 
These behaviors are understood by decomposing $\Lambda_{zxx}$ into the contribution for the $p$th band,  $\Lambda^{(p)}_{zxx}$ ($p=0$-$5$), i.e., $\Lambda_{zxx}=\sum_p \Lambda^{(p)}_{zxx}$.  
When the bands are well separated by $\pm \sqrt{2}h$ for large $h=5$ in Fig.~\ref{Fig:MultiOrbital_distortion}(c), two maxima arise from the lower ($p=0,1$) and higher ($p=4,5$) bands and one minimum arises from the middle band ($p=2,3$). 
With decrease of $h$, the separated bands become closer with each other, and then, they are overlapped at the band edge, as shown in Fig.~\ref{Fig:MultiOrbital_distortion}(b). 
With further decrease of $h$, the sum of the different band contributions results in the complicated behavior, as shown in Fig.~\ref{Fig:MultiOrbital_distortion}(a). 

To examine the behavior of $\Lambda_{zxx}$ in the MQ ordered state in detail, we compare it with the order parameter $\langle M^{\rm (h)}_{xy} \rangle$, plotted as a function of $\mu$ at $h=5$ in Fig.~\ref{Fig:MultiOrbital_distortion}(d).
$|\langle M^{\rm (h)}_{xy} \rangle|$ increases while decreasing $|\mu|$ and shows almost constant for $|\mu|<3.7$. 
This result indicates that the behaviors of $\langle M^{\rm (h)}_{xy} \rangle$ do not
have simple correlation like $\Lambda_{zxx} \propto \langle M^{\rm (h)}_{xy} \rangle$ except in the region for the low/high-electron density. 
Besides, we also compare $\Lambda_{zxx}$ with the $\mu$ derivative of $\langle M^{\rm (h)}_{xy} \rangle$, $\langle M^{\rm (h)}_{xy} \rangle_{E_{\rm F}}$, since $\Lambda_{zxx}$ is characterized by the intraband process at the Fermi surface. 
As shown in Fig.~\ref{Fig:MultiOrbital_distortion}(d), $\langle M^{\rm (h)}_{xy} \rangle_{E_{\rm F}}$ is enhanced at the inflection points with $|\mu| \simeq 7.1$, but it vanishes in region for $|\mu|<3.7$. 
Thus, $\Lambda_{zxx}$ do not have simple correlation with $\langle M^{\rm (h)}_{xy} \rangle_{E_{\rm F}}$ as well.

On the other hand, we find that $\Lambda_{zxx}$ has strong correlation with the antisymmetric spin-orbital polarization at the Fermi surface $-\langle \sin k_x l_z \sigma_x \rangle_{E_{\rm F}}$ when the Fermi surface has the simple form, as shown in Fig.~\ref{Fig:MultiOrbital_distortion}(d); there are two broad maxima at $|\mu| \simeq 7.1$ and the broad minimum at $\mu=0$. 
This result indicates that the quantity of $-\langle \sin k_x l_z \sigma_x \rangle_{E_{\rm F}}$, which is related with the spin-orbital momentum locking, becomes the appropriate measure of the current-induced distortion. 
In other words, a large response is expected when the degree of the spin-orbital momentum locking becomes large. 
There are two possibilities to reach a large value of $\Lambda_{zxx}$: 
One is the large value of $\mathcal{O}_{\mu}(\bm{k})$ at the Fermi surface and the other is the large density of states (DOS) denoted as the dotted lines in Fig.~\ref{Fig:MultiOrbital_distortion}(d). 
For the former $\mathcal{O}_{\mu}(\bm{k})$, a larger $s$-$p$ hopping is preferable as discussed in Sec.~\ref{sec:Band structure}. 
Meanwhile, for the latter, the large enhancement of the density of states, such as the van Hove singularity or flat band, is required. 
Indeed, $\Lambda_{zxx}$ shows an increase by approaching $\mu \simeq 0, \pm \sqrt{2}h$, where the small circular Fermi surface close to the band edge in Figs.~\ref{Fig:MultiOrbital_distortion_FS}(a) and \ref{Fig:MultiOrbital_distortion_FS}(c) gradually changes to the large square-shaped one at the van Hove singularity arising from the square-lattice geometry in Figs.~\ref{Fig:MultiOrbital_distortion_FS}(b) and \ref{Fig:MultiOrbital_distortion_FS}(d).

\subsubsection{Magnetoelectric effect}
\label{sec:Magnetoelectric effect}

\begin{figure}[htb!]
\begin{center}
\includegraphics[width=0.9 \hsize]{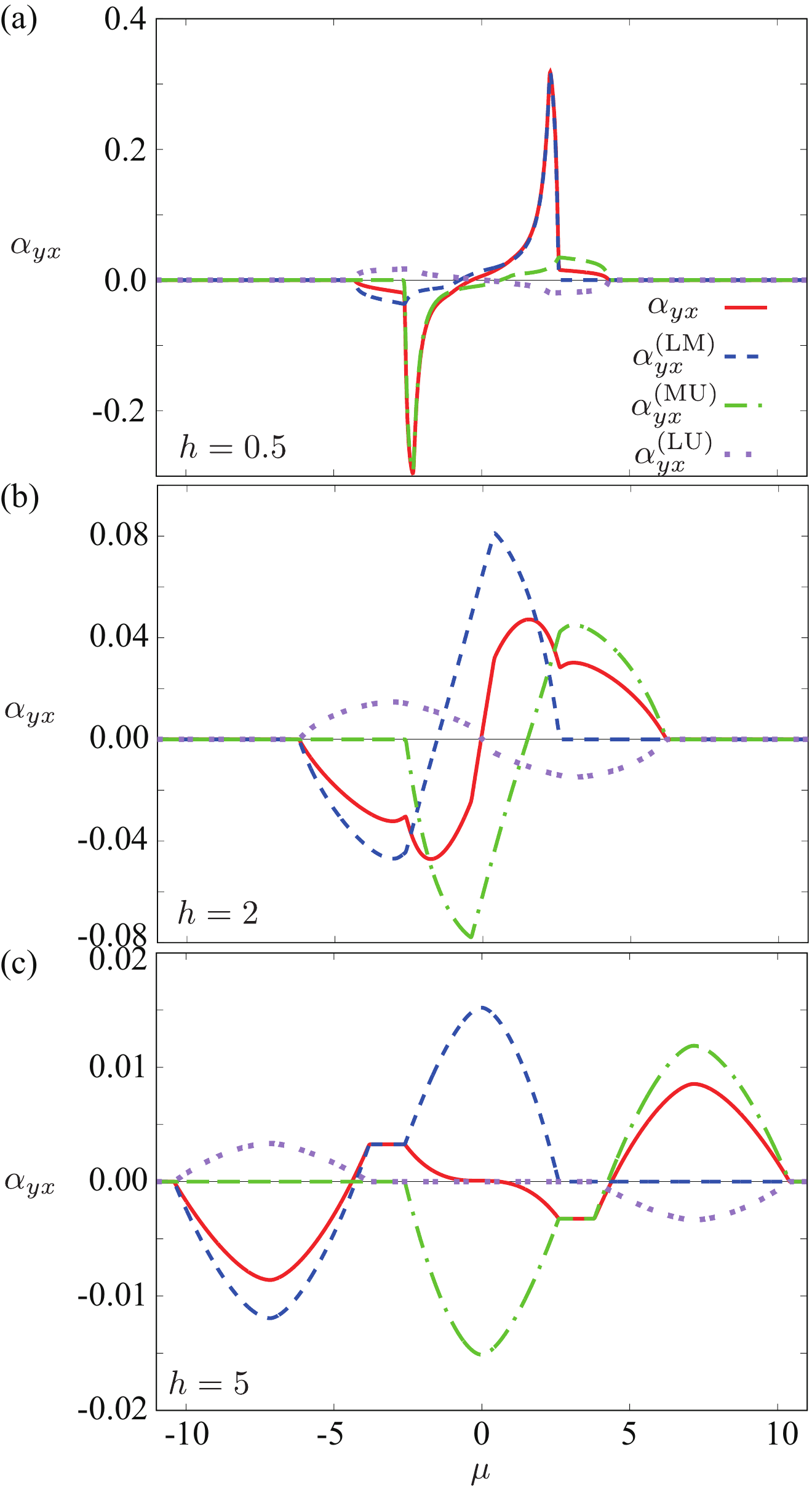} 
\caption{
\label{Fig:MultiOrbital_ME}
$\mu$ dependence of the coefficient of the current-magnetization correlation, $\alpha_{yx}$, in the $M^{\rm (h)}_{xy}$ state at temperature $T=0.01$ and the damping factor $\delta=0.01$ for (a) $h=0.5$, (b) $2$, and (c) $5$. 
$\alpha^{\rm (LM)}_{yx}$, $\alpha^{\rm (MU)}_{yx}$, and $\alpha^{\rm (LU)}_{yx}$ represents the interband contribution for the specific bands. 
L, M, and U represents the lower two bands ($p=0,1$), middle two bands ($p=2,3$), and upper two bands ($p=4,5$), respectively.
For example, $\alpha^{\rm (LM)}_{yx}$ stands for the interband process between the lower two bands and middle two bands. 
The other parameters are the same as Fig.~\ref{Fig:MultiOrbital_band}. 
}
\end{center}
\end{figure}

\begin{figure}[htb!]
\begin{center}
\includegraphics[width=0.8 \hsize]{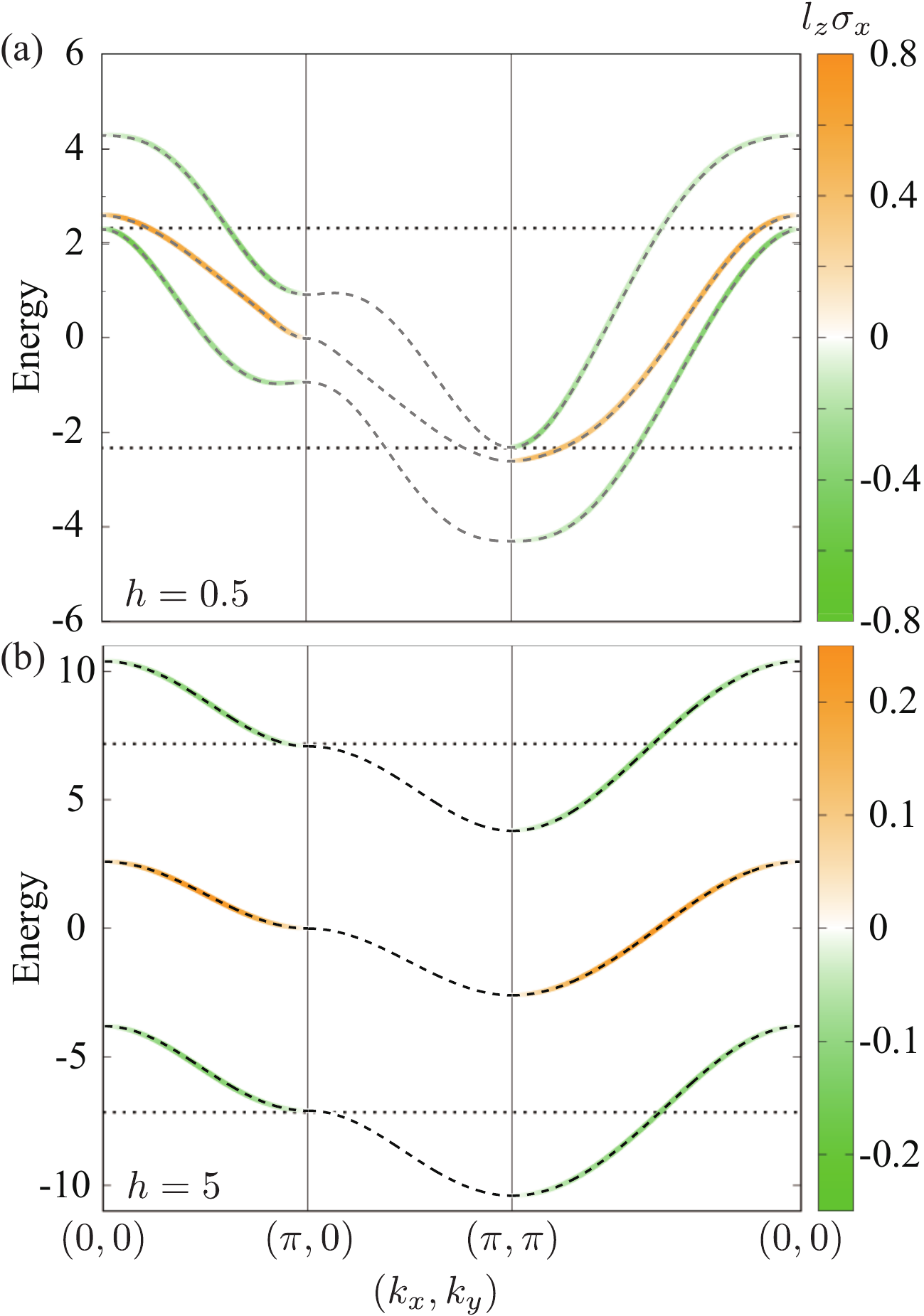} 
\caption{
\label{Fig:MultiOrbital_ME_band}
The band structure at (a) $h=0.5$ and (b) $5$. 
The dashed curves show the band dispersions and the colormap shows the spin-orbital polarization of the $l_z \sigma_x$ component at each wave vector. 
The horizontal dotted lines show the energy where $|\alpha_{yx}|$ is maximized. 
}
\end{center}
\end{figure}

Next, we consider another cross-correlated response in the MQ ordered state. 
We investigate the magnetoelectric effect where the magnetization $M_{\mu}$ is induced by the electric field $E_{\nu}$, i.e., $M_\mu = \sum_\nu \alpha_{\mu \nu} E_\nu$ for $\mu, \nu = x,y,z$. 
The tensor $\alpha_{\mu\nu}$ is calculated by the linear response theory as 
\begin{align}
\label{eq:ME}
\alpha_{\mu\nu} & = 
 \sum_{\bm k} \sum_{pq} \Pi_{pq}({\bm k}) M_{\mu{\bm k}}^{pq}v_{\nu{\bm k}}^{qp},
\end{align}
where $M^{pq}_{\mu {\bm k}}= \braket{p{\bm k}|\sigma_\mu|q{\bm k}}$ is the matrix element of the spin. 
We here take into account only the spin component in $M^{pq}_{\mu {\bm k}}$ for simplicity. 

Similar to the current-induced distortion tensor $\Lambda_{\zeta\nu}$ in previous section, the magnetoelectric tensor $\alpha_{\mu\nu}$ becomes nonzero in the absence of the spatial inversion symmetry. 
Since the time-reversal property between $\Lambda_{\zeta\nu}$ and $\alpha_{\mu\nu}$ is opposite, the odd-parity magnetic and magnetic toroidal multipoles with time-reversal odd contribute to the interband process $p \neq q$.  
By using the active rank 0-2 odd-parity multipoles, the magnetic monopole $M_0$, the magnetic toroidal dipole $\{ T_x, T_y, T_z \}$, and MQ $\{M_u, M_v, M_{yz}, M_{zx}, M_{xy}\}$, the matrix form of $\alpha_{\mu\nu}$ is represented by  
\begin{align}
&
\alpha=
\begin{pmatrix}
M_{0}-M_{u}+M_{v} & M_{xy}+T_{z} & M_{zx}-T_{y} \\
M_{xy}-T_{z} & M_{0}-M_{u}-M_{v} & M_{yz}+T_{x} \\
M_{zx}+T_{y} & M_{yz}-T_{x} & M_{0}+2M_{u}
\end{pmatrix}, 
\end{align}
where the row (column) of the matrix represents the component of $\{E_x, E_y, E_z\}$ ($\{M_x, M_y, M_z\}$). 
The nonzero matrix elements under the MQ ordering are shown in Table~\ref{tab_mp}. 
In the following, we focus on the behavior of $\alpha_{\mu\nu}$ in the $M^{\rm (h)}_{xy}$ ordered state with $\alpha_{xy}=\alpha_{yx}$. 

Figure~\ref{Fig:MultiOrbital_ME} shows the $\mu$ dependence of $\alpha_{yx}(=\alpha_{xy}$) denoted by the red solid lines in the $M^{\rm (h)}_{xy}$ ordered state for (a) $h=0.5$, (b) $h=2$, and (c) $h=5$. 
The other parameters are common in Sec.~\ref{sec:Band structure}. 
The result shows that $\alpha_{yx}$ is induced by nonzero $h$ as $\Lambda_{zxx}$. 
In contrast to $\Lambda_{zxx}$, $\alpha_{yx}$ takes a finite value in the insulating region for $2.7<|\mu|<3.7$ and $h=5$ in addition to the metallic region, since the interband process dominates in $\alpha_{yx}$. 
As compared with the results in Figs.~\ref{Fig:MultiOrbital_ME}(a)-\ref{Fig:MultiOrbital_ME}(c), $\alpha_{yx}$ tends to be smaller for larger $h$, which is reasonable in terms of the interband process: the larger energy difference in the denominator in Eq.~(\ref{eq:ME}) for larger $h$ suppresses $\alpha_{yx}$. 
Although nonzero $\alpha_{yx}$ exists in the presence of the antisymmetric spin-orbital polarization under the MQ ordering, its behavior is mainly determined by the details of the electronic band structure, as discussed below. 

As each band is doubly degenerate owing to the $\mathcal{PT}$ symmetry, the total six bands are separated into three two-degenerate bands under the MQ ordering. 
Then, $\alpha_{yx}$ is decomposed into three parts according to the different interband processes, $\alpha_{yx}=\alpha^{\rm (LM)}_{yx}+\alpha^{\rm (MU)}_{yx}+\alpha^{\rm (LU)}_{yx}$ where the superscripts L, M, and U represent the lower two bands $p=0,1$, middle two bands $p=2,3$, and upper two bands $p=4,5$, respectively. 
In other words, $\alpha^{\rm (LM)}_{yx}$ includes the contribution of the interband process between the lower two bands and middle two bands, for instance. 
For small $h=0.5$, the contribution of $\alpha^{\rm (MU)}_{yx}$ ($\alpha^{\rm (LM)}_{yx}$) is dominant for the negative (positive) peak at $\mu \simeq -2.33$ (2.32), while $\alpha^{\rm (LU)}_{yx}$ is less important. 
The large enhancement at $\mu \simeq -2.33$ and $2.32$ is attributed to the small band gap in the electronic band structure for small $h$, as shown in Fig.~\ref{Fig:MultiOrbital_ME_band}(a). 
The band structure in Fig.~\ref{Fig:MultiOrbital_ME_band}(a) indicates that the dominant contribution comes from near the $\bm{k}=(\pi,\pi)$ [$\bm{k}=(0,0)$] point at $\mu \simeq -2.33$ (2.32), which is originally fourfold degenerate at $h=0$. 
When the three bands are separated by increasing $h$, the contribution of $\alpha^{\rm (LM)}_{yx}$ ($\alpha^{\rm (MU)}_{yx}$) becomes important for low (high) electron density, as shown in Figs.~\ref{Fig:MultiOrbital_ME}(b) and \ref{Fig:MultiOrbital_ME}(c). 
For large $h$, all the $\bm{k}$ points below the Fermi level contribute to $\alpha^{\rm (LM)}_{yx}$ irrespective of $\bm{\mathcal{Q}}(\bm{k})$, since the energy difference between the lower and middle bands at each $\bm{k}$ takes similar values, as shown in Fig.~\ref{Fig:MultiOrbital_ME_band}(b). 
The broad peaks at $\mu \simeq -7.16$ and $7.17$ for $h=5$ are attributed to the van Hove singularity, as shown in Fig.~\ref{Fig:MultiOrbital_distortion_FS}(b).
Meanwhile, $\alpha_{yx}$ is negligibly small close to the half filling owing to the cancellation of the contributions of $\alpha^{\rm (LM)}_{yx}$ and $\alpha^{\rm (MU)}_{yx}$, as shown in Fig.~\ref{Fig:MultiOrbital_ME}(c).

\subsection{Sublattice system}
\label{sec:Multi-sublattice system}

Next, we investigate another situation with the active MQ degree of freedom in the sublattice system. 
We consider a four-sublattice model in the tetragonal system in Sec.~\ref{sec:Model_2}. 
In Sec.~\ref{sec:Band structure_2}, we show that a similar spin-orbital momentum locking occurs in the MQ ordering even without the atomic orbital degree of freedom.

\subsubsection{Model}
\label{sec:Model_2}

\begin{figure}[htb!]
\begin{center}
\includegraphics[width=1.0 \hsize]{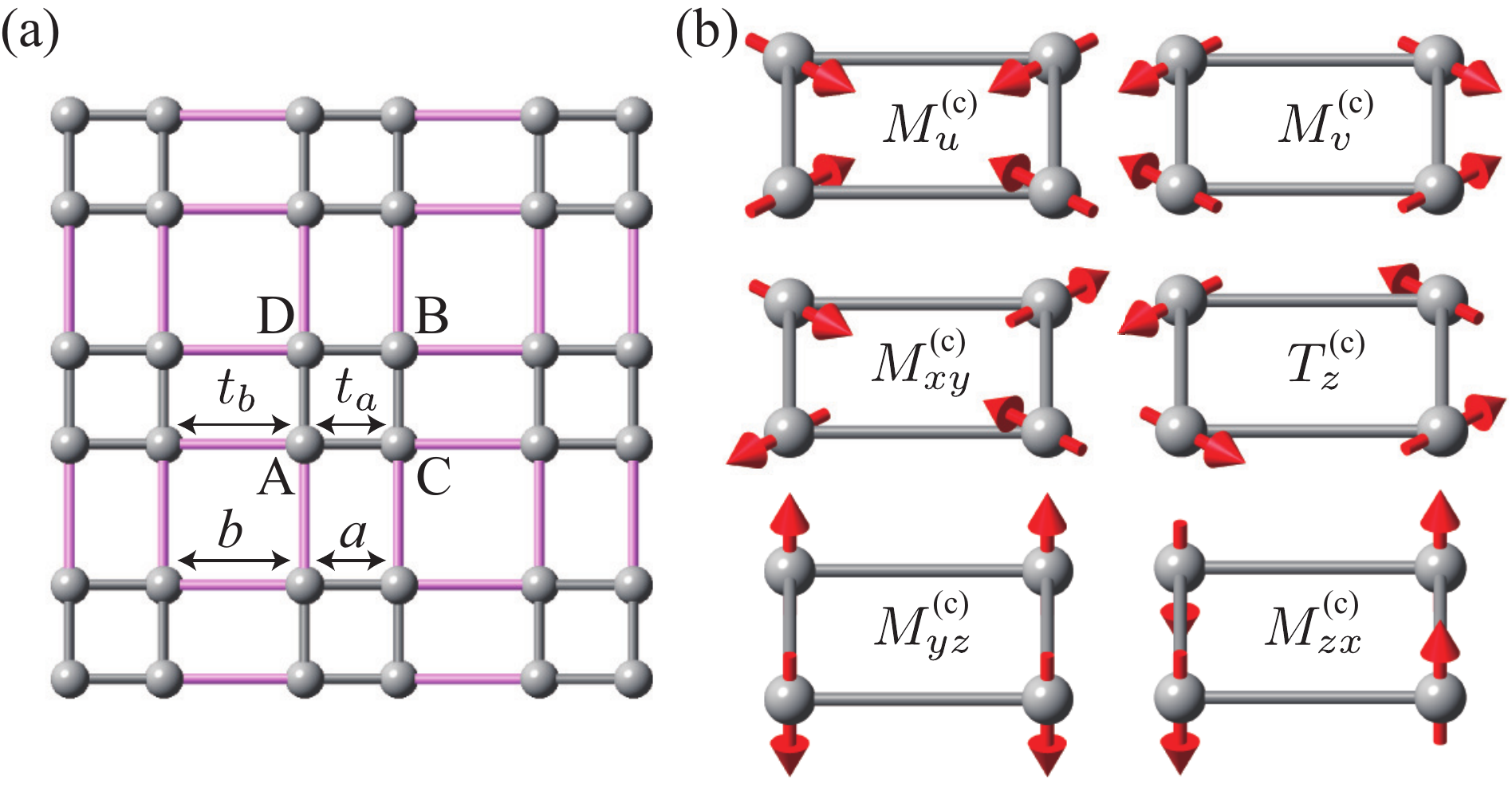} 
\caption{
\label{Fig:Multisite}
(a) Sublattice model consisting of four sublattices A-D in the tetragonal lattice structure with the lattice constant $a+b$. 
(b) The spin patterns of the odd-parity cluster MQs $\{M^{\rm (c)}_u, M^{\rm (c)}_v, M^{\rm (c)}_{yz}, M^{\rm (c)}_{zx}, M^{\rm (c)}_{xy}\}$ and magnetic toroidal dipole $T^{\rm (c)}_z$ in Eqs.~(\ref{eq:Mu_multisite})-(\ref{eq:Tz_multisite}).
The arrow represents the direction of magnetic moments. 
}
\end{center}
\end{figure}

In this section, we consider another electronic degree of freedom to activate the MQ. 
In particular, we focus on the sublattice degree of freedom instead of the orbital one, where the MQ is activated with the antiferromagnetic ordering.  
We examine the four-sublattice system in the tetragonal lattice structure, as shown in Fig.~\ref{Fig:Multisite}(a), where the point group is $D_{4 {\rm h}}$ as in Sec.~\ref{sec:Multi-orbital system}. 
The lattice constant is taken as $a=b=1/2$ for notational simplicity and the difference between $a$ and $b$ is expressed as the different hopping amplitudes, $t_a$ and $t_b$. 
When there is no orbital degree of freedom at each sublattice, the tight-binding Hamiltonian is given by 
\begin{align}
\label{eq:Ham_multisite}
\mathcal{H}=\sum_{\bm{k},\gamma,\gamma',\sigma}c^{\dagger}_{\bm{k}\gamma \sigma} H^{\gamma\gamma'}_\sigma c_{\bm{k}\gamma'\sigma}, 
\end{align}
where $c^{\dagger}_{\bm{k}\gamma \sigma}$ ($c_{\bm{k}\gamma \sigma}$) is the creation (annihilation) operator of electrons at wave vector $\bm{k}$, sublattice $\gamma=$A-D, and spin $\sigma$. 
The $4\times 4$ Hamiltonian matrix spanned by the four-sublattice basis $\{ \phi_{{\rm A}\sigma}, \phi_{{\rm B}\sigma}, \phi_{{\rm C}\sigma}, \phi_{{\rm D}\sigma} \}$ is given by 
\begin{align}
\label{eq:Ham_4sub}
H_{\sigma}=\left(
\begin{array}{cccccc}
0 & 0 & f^*_x & f^*_y\\
0 & 0 & f_y & f_x\\
f_x & f^*_y & 0 & 0\\
f_y & f^*_x & 0 & 0
\end{array}
\right),  
\end{align}
where $f_{\eta} = t_a e^{i k_{\eta}/2}+t_b e^{-i k_{\eta}/2}$ for $\eta=x,y$. 

The Hamiltonian in Eq.~(\ref{eq:Ham_multisite}) has 64 independent electronic degrees of freedom. 
Similar to the discussion in Sec.~\ref{sec:Multi-orbital system}, one can construct the MQ degree of freedom by the product of the odd-parity electronic degree of freedom in spinless space and spin $\sigma$. 
The spinless odd-parity electronic degree of freedom is expressed as the spatial distribution of the onsite potential with the same magnitude but the different sign, which corresponds to the odd-parity electric dipoles, $Q^{\rm (c)}_x$ and $Q^{\rm (c)}_y$. 
The matrix forms of $Q^{\rm (c)}_x$ and $Q^{\rm (c)}_y$ are given by 
\begin{align}
Q^{\rm (c)}_x&=
\left(
\begin{array}{cccc}
-1 & 0 & 0&0 \\
0 & 1 & 0 &0\\
0 & 0 & 1 &0\\
0 & 0 & 0 & -1 
\end{array}
\right), \\
Q^{\rm (c)}_y&=
\left(
\begin{array}{cccc}
-1 & 0 & 0&0 \\
0 & 1 & 0 &0\\
0 & 0 & -1 &0\\
0 & 0 & 0 & 1 
\end{array}
\right), 
\end{align}
where the superscript (c) means the cluster multipole that is defined in the sublattice cluster~\cite{hayami2016emergent,Suzuki_PhysRevB.95.094406,Suzuki_PhysRevB.99.174407}. 
We here do not explicitly consider the other electric dipole degree of freedom, e.g., the bond degree of freedom, and we here focus on the antiferromagnetic ordering, which is represented by the contraction of $\{Q^{\rm (c)}_x, Q^{\rm (c)}_y\}$ and $\bm{\sigma}$. 
By taking a linear combination of them, we obtain the expressions for five MQs as
\begin{align}
\label{eq:Mu_multisite}
A_{1u}^{-}:\quad M^{\rm (c)}_u &= -Q^{\rm (c)}_x \sigma_x -Q^{\rm (c)}_y \sigma_y, \\
\label{eq:Mv_multisite}
B_{1u}^{-}:\quad M^{\rm (c)}_v &= Q^{\rm (c)}_x \sigma_x - Q^{\rm (c)}_y \sigma_y, \\
\label{eq:Myz_multisite}
E_{u}^{-}:\quad M^{\rm (c)}_{yz} &= Q^{\rm (c)}_y \sigma_z, \\
\label{eq:Mzx_multisite}
M^{\rm (c)}_{zx} &= Q^{\rm (c)}_x \sigma_z, \\
\label{eq:Mxy_multisite}
B_{2u}^{-}:\quad  M^{\rm (c)}_{xy} &= Q^{\rm (c)}_x \sigma_y + Q^{\rm (c)}_y \sigma_x, 
\end{align}
and the magnetic toroidal dipole as 
\begin{align}
\label{eq:Tz_multisite}
A_{2u}^{-}:\quad T^{\rm (c)}_{z} &= Q^{\rm (c)}_x \sigma_y - Q^{\rm (c)}_y \sigma_x. 
\end{align}
The corresponding spin patterns are shown in Fig.~\ref{Fig:Multisite}(b), where $M^{\rm (c)}_u$, $M^{\rm (c)}_v$, $M^{\rm (c)}_{xy}$, and $T^{\rm (c)}_z$ exhibit the noncollinear magnetic textures, while $M^{\rm (c)}_{yz}$ and $M^{\rm (c)}_{zx}$ are the collinear magnetic textures. 
It is noted that $M^{\rm (c)}_{yz}$ and $M^{\rm (c)}_{zx}$ are also regarded as the magnetic toroidal dipole ordering $T^{\rm (c)}_{x}$ and $T^{\rm (c)}_{y}$, respectively, which belong to the same irreducible representation under the point group $D_{4{\rm h}}$.

\subsubsection{Band structure}
\label{sec:Band structure_2}

\begin{figure}[htb!]
\begin{center}
\includegraphics[width=1.0 \hsize]{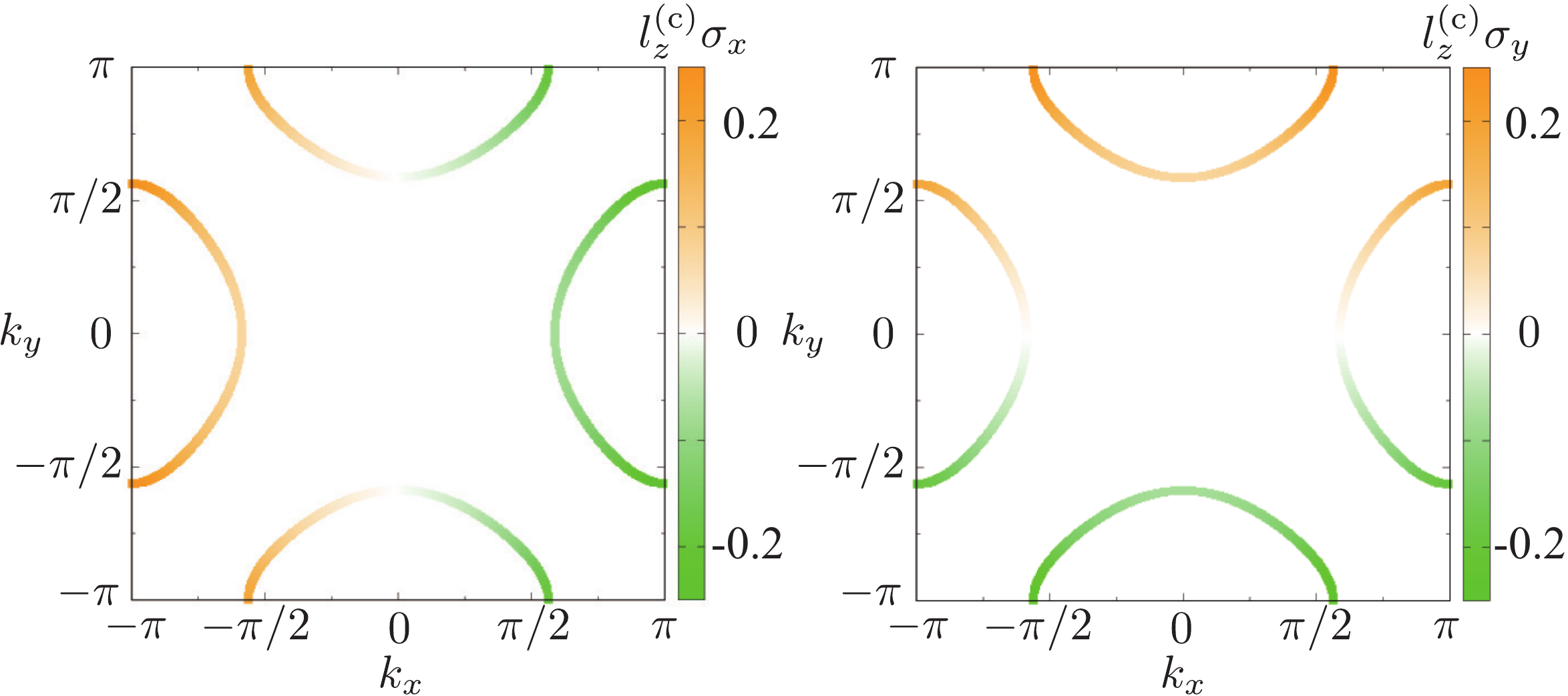} 
\caption{
\label{Fig:Multisite_band}
The isoenergy surfaces at $\mu=-0.5$, $t_a=1$, $t_b=0.5$, and $h=0.3$ in the $M_{xy}$ state. 
The colormap shows the spin-orbital polarization of the $l^{\rm (c)}_z \sigma_x$ (left) and $l^{\rm (c)}_z \sigma_y$ (right) at each wave vector.   
}
\end{center}
\end{figure}

Since there is no orbital degree of freedom in the system, orbital angular momentum $\bm{l}$ is inactive as the electronic degree of freedom in the Hamiltonian. 
Nevertheless, a similar pseudo-orbital angular momentum is defined as the degree of freedom over the sublattices. 
Since the electron hoppings on the closed loop in the square plaquette as A $\to$ C $\to$ B $\to$ D give rise to the magnetic flux to the electrons in the out-of-plane direction, the matrix of the pseudo-orbital angular momentum is defined as 
\begin{align}
l^{\rm (c)}_z&=
\left(
\begin{array}{cccc}
0 & 0 & -i & i \\
0 & 0 & i & -i\\
i & -i & 0 &0\\
-i & i & 0 & 0 
\end{array}
\right). 
\end{align}
By using $l^{\rm (c)}_z$ instead of $l_z$ in Eqs.~(\ref{eq:Muk_orbital})-(\ref{eq:Mxyk_orbital}), one can obtain similar physics in Sec.~\ref{sec:Multi-orbital system}. 

We show a similar spin-orbital momentum locking in the model in Eq.~(\ref{eq:Ham_multisite}) under the MQ ordering. 
We here focus on the $M_{xy}$ ordered state in Fig.~\ref{Fig:Multisite}(b) with the molecular field $h$ as an example. 
Figure~\ref{Fig:Multisite_band} shows the isoenergy surfaces at $\mu=-0.5$, $t_a=1$, $t_b=0.5$, and $h=0.3$ in the $M_{xy}$ ordered state. 
Similar to the result in Fig.~\ref{Fig:MultiOrbital_band}(c), the band structure exhibits the antisymmetric spin-orbital polarization of $l^{\rm (c)}_z \sigma_x$ and $l^{\rm (c)}_z \sigma_y$, which corresponds to the spin-orbital momentum locking. 
The component of $l^{\rm (c)}_z \sigma_x$ is asymmetric with respect to the $k_x$ direction, while that of $l^{\rm (c)}_z \sigma_y$ is asymmetric with respect to the $k_y$ direction. 
The necessary hopping parameters are also obtained by evaluating $\mathcal{O}_{\mu}(\bm{k}) ={\rm Tr}[e^{-\beta \mathcal{H}_{\bm{k}}} l^{\rm (c)}_z \sigma_\mu]$. 
The lowest-order contributions are given by 
\begin{align}
-h(t_a-t_b)\left(\sin \frac{k_x}{2} l^{\rm (c)}_z \sigma_x + \sin \frac{k_y}{2} l^{\rm (c)}_z \sigma_y\right),  
\end{align}
which indicates that the relation $t_a \ne t_b$ is necessary for the spin-orbital momentum locking besides $h \ne 0$. 
Also in this case, the effective spin-orbit coupling in the form of $l^{\rm (c)}_z \sigma_z$ emerges even without the effect of the atomic relativistic spin-orbit coupling in the tight-binding model (not shown). 
The contribution is represented by $h^2 (t_a + t_b)(\cos k_x/2 + \cos k_y/2)$

Once the spin-orbital momentum locking occurs by the MQ ordering, similar physics discussed in Sec.~\ref{sec:Multi-orbital system} is expected, such as the antisymmetric spin polarization by the magnetic field and linear responses. 
We briefly discuss the electronic band structure in the presence of the magnetic field. 
Although similar symmetric/antisymmetric spin splittings shown in Fig.~\ref{Fig:MultiOrbital_band_H} are expected because of the same symmetry, the necessary model parameters are different from the multi-orbital case. 
When the magnetic field is applied along the $x$ direction, the $k_x k_y \sigma_y$-type symmetric and $k_x \sigma_z$-type antisymmetric spin splittings are expected, but they do not show up within the model Hamiltonian in Eq.~(\ref{eq:Ham_multisite}). 
Such spin splittings under the magnetic field appear when the additional diagonal hopping between A and B (and C and D) and/or the effective spin-dependent hopping from the atomic relativistic spin-orbit coupling exist. 
Considering such terms within the plaquette, we obtain the former 
Hamiltonian matrix as 
\begin{align}
\label{eq:Ham_diaghop}
H^{\rm diag}_{\sigma}=\left(
\begin{array}{cccccc}
0 & f^*_{xy} & 0 & 0\\
f_{xy} & 0 & 0 & 0\\
0 & 0 & 0 & f'^*_{xy}\\
0 & 0 & f'_{xy} & 0
\end{array}
\right),  
\end{align}
where $f_{xy}=t'_a e^{i(k_x+k_y)/2}$ and $f'_{xy}=t'_a e^{i(-k_x+k_y)/2}$ and the latter as 
\begin{align}
\label{eq:Ham_SOC_cluster}
H^{\rm SOC}_{\sigma}=\lambda \left(
\begin{array}{cccccc}
0 & 0 & -i e^{-i k_x/2} & i e^{-i k_y/2}\\
0 & 0 & i e^{i k_y/2} & -i e^{i k_x/2}\\
i e^{i k_x/2} & -i e^{-i k_y/2} & 0 & 0\\
-i e^{i k_y/2} & i e^{-i k_x/2} & 0 & 0
\end{array}
\right).  
\end{align}

By evaluating $\mathcal{O}_{\mu}(\bm{k})$, we obtain the necessary effective coupling for the symmetric spin splitting $k_x k_y \sigma_y$ as $h^2 H_x t^2_b t'_a \sin k_x \sin k_y $ and the antisymmetric spin splitting $k_x \sigma_z$ as the superposition of $h H_x \lambda t_b \sin k_x$ and $h^3 H_x t_a t'_a t_b \sin k_x$. 
Thus, $t'_a$ is necessary for both the symmetric and antisymmetric spin splittings, while $\lambda$ is the antisymmetric spin splitting. 
In a similar way, the antisymmetric spin splitting $k_x \sigma_x - k_y \sigma_y$ in a magnetic field along $z$-axis is caused by introducing $t'_a$ and $\lambda$, e.g., the coupling $-h H_z \lambda t_b \sin k_x \sigma_x$ and $h^3 H_z t_a t'_a t_b \sin k_x \sigma_x$.

\section{Summary}
\label{sec:Summary}

To summarize, we have investigated the electronic states and related physical phenomena induced by the MQ orderings. 
We found that the MQ ordered state exhibits a peculiar spin-orbital entanglement in momentum space; the spin-orbital polarization is antisymmetrically locked at the particular component at each wave vector, which is dubbed the spin-orbital momentum locking. 
The present spin-orbital momentum locking is driven by the onset of the MQ orderings in contrast to the spin momentum locking that exists in the nonmagnetic noncentrosymmetric lattice systems. 
We show typical two examples for the MQ orderings by considering the multi-orbital and sublattice systems. 
We demonstrate that the spin-orbital momentum locking occurs under the MQ orderings, which causes various cross-correlated physical phenomena, such as the magnetic-field-induced symmetric and antisymmetric spin polarization in the band structure, the current-induced distortion, and the magnetoelectric effect. 
We discuss the relevant model parameters in each phenomenon. 
As the spin-orbital momentum locking is expected to be found in odd-parity magnetic materials not only in the MQ phase but also in other magnetic and magnetic toroidal multipole phases, our study will stimulate a further exploration of functional spintronics materials, which have recently been extensively studied.

\begin{acknowledgments}
This research was supported by JSPS KAKENHI Grants Numbers JP15H05885, JP18H04296 (J-Physics), JP18K13488, JP19K03752, JP19H01834, JP21H01037, and by JST PREST (JPMJPR20L8). 
Parts of the numerical calculations were performed in the supercomputing systems in ISSP, the University of Tokyo, and in the MAterial science Supercomputing system for Advanced MUlti-scale simulations towards NExt-generation-Institute for Materials Research (MASAMUNE-IMR) of the Center for Computational Materials Science, Institute for Materials Research, Tohoku University.
\end{acknowledgments}

\bibliographystyle{apsrev}
\bibliography{ref}

\end{document}